\documentclass[fleqn,usenatbib]{mnras}

\usepackage[T1]{fontenc}
\usepackage{ae,aecompl}
\usepackage{graphicx}	
\usepackage{amsmath}	
\usepackage{amssymb}	

\newcommand{\cs}{c_\mathrm{s}}
\newcommand{\pc}{\mathrm{pc}}
\newcommand{\kpc}{\mathrm{kpc}}
\newcommand{\yr}{\mathrm{yr}}
\newcommand{\Myr}{\mathrm{Myr}}
\newcommand{\s}{\mathrm{s}}
\newcommand{\km}{\mathrm{km}}
\newcommand{\msol}{\mathrm{M_{\odot}}}
\newcommand{\tff}{t_\mathrm{ff}}
\newcommand{\err}{\mathrm{err}}
\newcommand{\gastff}{(\siggas/t)_\mathrm{single-ff}}
\newcommand{\gastmff}{(\siggas/t)_\mathrm{multi-ff}}
\newcommand{\mach}{\mathcal{M}}

\newcommand{\sigsfr}{\Sigma_\mathrm{SFR}}
\newcommand{\siggas}{\Sigma_\mathrm{gas}}
\newcommand{\sigsfrunit}{\mathrm{\msol\,\yr^{-1}\,\kpc^{-2}}}
\newcommand{\siggasunit}{\mathrm{\msol\,\pc^{-2}}}

\newcommand{\cas}{Composite/AGN/Shock}

\title[The SAMI Galaxy Survey: estimating molecular gas]
{The SAMI Galaxy Survey: a new method to estimate molecular gas surface densities from star formation rates}

\author[Federrath et al.]
{Christoph~Federrath$^1$, Diane~M.~Salim$^{1,2}$, Anne~M.~Medling$^{1,3,4}$, 
\newauthor{Rebecca~L.~Davies$^{1,5}$, Tiantian~Yuan$^1$, Fuyan~Bian$^1$, Brent~A.~Groves$^1$,}   
\newauthor{I-Ting~Ho$^{6,1}$, Robert~Sharp$^1$, Lisa J.~Kewley$^1$, Sarah~M.~Sweet$^1$,} 
\newauthor{Samuel~N.~Richards$^{7,8,2}$, Julia J. Bryant$^{7,8,2}$, Sarah~Brough$^7$, Scott~Croom$^{8,2}$,}
\newauthor{Nicholas Scott$^{8,2}$, Jon Lawrence$^7$, Iraklis Konstantopoulos$^7$, Michael Goodwin$^7$}\\\\
$^1$Research School of Astronomy and Astrophysics, Australian National University, Canberra, ACT 2611, Australia \\
$^2$ARC Centre of Excellence for All-sky Astrophysics (CAASTRO) \\
$^3$Cahill Center for Astronomy and Astrophysics, California Institute of Technology, MS 249-17, Pasadena, CA 91125, USA \\
$^4$Hubble Fellow \\
$^5$Max-Planck-Institut f\"ur Extraterrestrische Physik, Giessenbachstrasse, D-85748 Garching, Germany \\
$^6$Institute for Astronomy, University of Hawaii, 2680 Woodlawn Drive, Honolulu, HI 96822, USA\\
$^7$Australian Astronomical Observatory, PO~Box~915, North Ryde, NSW~1670, Australia \\
$^8$Sydney Institute for Astronomy, School of Physics, University of Sydney, NSW~2006, Australia \\
\vspace{-0.3cm}\\For enquiries please email the corresponding author: \href{mailto:christoph.federrath@anu.edu.au}{christoph.federrath@anu.edu.au}
}

\pagerange{\pageref{firstpage}--\pageref{lastpage}} \pubyear{2017}

\begin{document}

\label{firstpage}

\maketitle

\begin{abstract}
Stars form in cold molecular clouds. However, molecular gas is difficult to observe because the most abundant molecule ($\mathrm{H_2}$) lacks a permanent dipole moment. Rotational transitions of CO are often used as a tracer of $\mathrm{H_2}$, but CO is much less abundant and the conversion from CO intensity to $\mathrm{H_2}$ mass is often highly uncertain. Here we present a new method for estimating the column density of cold molecular gas ($\siggas$) using optical spectroscopy. We utilise the spatially resolved H$\alpha$ maps of flux and velocity dispersion from the Sydney-AAO Multi-object Integral-field spectrograph (SAMI) Galaxy Survey. We derive maps of $\siggas$ by inverting the multi-freefall star formation relation, which connects the star formation rate surface density ($\sigsfr$) with $\siggas$ and the turbulent Mach number ($\mach$). Based on the measured range of \mbox{$\sigsfr=0.005$--$1.5\,\sigsfrunit$} and \mbox{$\mach=18$--$130$}, we predict \mbox{$\siggas=7$--$200\,\siggasunit$} in the star-forming regions of our sample of 260 SAMI galaxies. These values are close to previously measured $\siggas$ obtained directly with unresolved CO observations of similar galaxies at low redshift. We classify each galaxy in our sample as `Star-forming' (219) or `{\cas}' (41), and find that in `{\cas}' galaxies the average $\sigsfr$, $\mach$, and $\siggas$ are enhanced by factors of $2.0$, $1.6$, and $1.3$, respectively, compared to Star-forming galaxies. We compare our predictions of $\siggas$ with those obtained by inverting the Kennicutt-Schmidt relation and find that our new method is a factor of two more accurate in predicting $\siggas$, with an average deviation of 32\% from the actual $\siggas$.
\end{abstract}



\section{Introduction}

The coalescence of gases by turbulence and gravity intricately controls star formation within giant molecular clouds \citep{Ferriere:2001aa,MacLowKlessen2004,ElmegreenScalo2004,ScaloElmegreen2004,McKeeOstriker2007,HennebelleFalgarone2012,Krumholz2014,PadoanEtAl2014}. On one hand, turbulence has the ability to hinder star formation by providing kinetic energy that can oppose gravity. On the other, the supersonic turbulence ubiquitously observed in the molecular phase of the interstellar medium (ISM) produces local shocks and compressions, which lead to enhanced gas densities that are key for triggering star formation \citep{FederrathKlessen2012}. Understanding the complex effects of turbulence in the ISM is therefore crucial to understanding the process of galaxy evolution.

The cold turbulent gas that provides the fuel for star formation is only visible in the millimetre/sub-millimetre to radio wavelengths, and is often faint, making it difficult to detect at high spatial resolutions. A standard method to measure the mean column density of molecular gas ($\siggas$) is to use rotational lines of CO. A severe problem with this method is that, because CO is about $10^{4}$ times less abundant than the main mass carrier, $\mathrm{H_2}$, one requires a CO-to-$\mathrm{H_2}$ conversion factor, which is typically calibrated based on measurements in our own galaxy. However, the CO-to-$\mathrm{H_2}$ conversion factor may depend on metallicity, environment and redshift, introducing high uncertainties in the reconstruction of the total gas surface densities from measurements of CO \citep{ShettyEtAl2011a,ShettyEtAl2011b}. Another method is to measure dust emission or dust extinction and assuming a gas-to-dust ratio to infer the molecular gas masses and surface densities. These methods can suffer from uncertainties in the gas-to-dust ratio, especially for low-metallicity galaxies where this ratio becomes increasingly uncertain. Both CO and dust observations require telescopes and instruments that work at millimetre/sub-millimetre wavelengths, which may not always be available and/or may have relatively low spatial resolution. Here we present a new method to estimate $\siggas$ based on the star formation rate (SFR), which can be obtained with optical spectroscopy.

Large optical integral field spectroscopy (IFS) surveys have started to provide us with details regarding the chemical distribution and kinematics of extragalactic sources at a size and uniformity unprecedented until recent times. Large galaxy surveys such as the Sloan Digital Sky Survey \citep[SDSS;][]{York:2000aa,Abazajian:2009aa}, 2-degreeField Galaxy Redshift Survey \citep[2dFGRS;][]{Colless:2001aa}, the Cosmic Evolution Survey \citep[COSMOS;][]{Scoville:2007aa}, the VIMOS VLT Deep Survey \citep[VVDS;][]{Le-Fevre:2004aa}, and the Galaxy and Mass Assembly survey \citep[GAMA;][]{Driver:2009aa,DriverEtAl2011} have contributed more than 3.5 million spectra that have been of extraordinary aid to our understanding of galaxy evolution. However, those spectra have been taken with a single fibre or slit, and provide only a single, global spectrum per galaxy \citep{BryantEtAl2015}. These spectra are therefore susceptible to aperture effects because differing parts or fractions of the galaxies are recorded for each source, thus making each observation dependent on the size and distance of the galaxy, as well as the positioning of the fibre \citep{RichardsEtAl2016}. Conversely, IFS can spatially resolve each galaxy observed, thus assigning individual spectra at many locations across the galaxy.  Here we utilise data from the SAMI galaxy survey, an IFS survey with the aim to observe 3400 galaxies over a broad range of environments and stellar masses. We use the SFRs measured in SAMI in order to provide a tool for estimating $\siggas$.

The basis of our $\siggas$ reconstruction method is a recent star formation relation developed in the multi-freefall framework of turbulent gas \citep{HennebelleChabrier2011,FederrathKlessen2012,Federrath2013sflaw,Salim:2015aa}. There have been many ongoing efforts to find an intrinsic relation between the amount of gas and the rate at which stars form in a molecular cloud. Initiated by \citet[][]{K98}, hereafter K98, $\sigsfr$ correlates with $\siggas$ \citep{Schmidt1959,K98,BigielEtAl2008,LeroyEtAl2008,Daddi:2010aa,SchrubaEtAl2011,RenaudEtAl2012,KennicuttEvans2012}, which can be approximated by an empirical power law with exponent $n$,
\begin{align} \label{eq:k98}
\sigsfr \propto \siggas^n.
\end{align}
For a sample of low-redshift disc and starburst galaxies, K98 found an exponent of $n=1.40\pm0.15$.
However, significant scatter and discrepancies between different sets of data exist within this framework, commonly referred to as the Kennicutt-Schmidt relation. These discrepancies suggest that $\sigsfr$ does not only depend on $\siggas$, but also on factors such as the turbulence and the freefall time of the dense gas on small scales.

Motivated by the fact that dense gas forms stars at a higher rate, a new star formation correlator was derived in \citet[][]{Salim:2015aa}, hereafter SFK15. This descriptor, denoted by $\gastmff$ and called the `maximum or multi-freefall gas consumption rate' (MGCR), is dependent on the probability density function \citep[PDF;][]{Vazquez1994,PadoanNordlundJones1997,PassotVazquez1998,FederrathKlessenSchmidt2008} of molecular gas,
\begin{align} \label{eq:newsflaw}
\sigsfr & = 0.45\% \times \gastmff \\
           & = 0.45\% \times \gastff \times \left( 1+b^2\mathcal{M}^2\frac{\beta}{\beta+1} \right)^{3/8},\nonumber
\end{align}
where $\mach$ is the Mach number of the turbulence, $b$ is the turbulence driving parameter \citep{FederrathKlessenSchmidt2008,Federrath:2010aa,FederrathEtAl2017iaus} and $\beta$ is the ratio of thermal to magnetic pressure \citep{PadoanNordlund2011,MolinaEtAl2012} in the molecular gas.

The SFK15 model for $\sigsfr$ given by Eq.~(\ref{eq:newsflaw}) is built upon foundational concepts laid out by \citet[][]{KDM12}, hereafter KDM12, which had parameterised $\sigsfr$ by the ratio between $\siggas$ and the average (single) freefall time $\tff$, a correlator hereon denoted by $\gastff$ \citep{KDM12,Federrath2013sflaw,Krumholz2014}. Our new correlator instead uses the concept of a multi-freefall time, which was pioneered by \citet{HennebelleChabrier2011}, tested with numerical simulations in \citet{FederrathKlessen2012}, and used in SFK15 as a stepping stone to expand upon the KDM12 model. SFK15 found that $\sigsfr$ is equal to $0.45\%$ of the MGCR by placing observations of Milky Way clouds and the Small Magellanic Cloud (SMC) in the K98, KDM12 and SFK15 frameworks, confirming the measured low efficiency of star formation \citep{KrumholzTan2007,Federrath2015}. Statistical tests in SFK15 showed that a significantly better correlation between $\sigsfr$ and $\gastmff$ was achieved than that which could be attained between either the $\siggas$ or $\gastff$ parameterisations of the previous star formation relations by K98 and KDM12, respectively. The scatter in the SFK15 relation was found to be a factor of 3--4 lower than in the K98 and KDM12 relations, suggesting that it provides a better physical model for $\sigsfr$ compared to the empirical relation by K98 and compared to the single-freefall relation by KDM12.

The aim of the current work is to formulate a method to predict the distribution of $\siggas$ by inverting Eq.~(\ref{eq:newsflaw}) and using optical observations, which will be plentiful in the coming few years. Here we use the H$\alpha$ luminosities and velocity dispersions provided by the SAMI Galaxy Survey to estimate $\siggas$ from measurements of $\sigsfr$ and $\mach$.

In Sec.~\ref{sec:obs} we describe the observations of our SAMI galaxy sample. Sec.~\ref{sec:theory} introduces our new method to derive $\siggas$ by inverting the SFK15 relation. In Sec.~\ref{sec:results} we present our results and compare purely star-forming with {\cas} galaxies in our sample. In Sec.~\ref{sec:comp} we compare our own and other observations and predictions to previous star formation relations within the Kennicutt-Schmidt framework. In Sec.~\ref{sec:k98vssfk15} we demonstrate that our new method for predicting $\siggas$ is superior to inverting the K98 relation. Our conclusions are summarised in Sec.~\ref{sec:conc}. The new data products for each SAMI galaxy in our sample derived here (average turbulent Mach number, cold gas density, freefall time, etc., and finally $\siggas$) are listed in Tab.~\ref{tab:galaxies} in Appendix~\ref{app:online} and are available for download in the online version of the journal or by contacting the authors.


\section{Sample Selection} \label{sec:obs}

\subsection{The SAMI Galaxy Survey} \label{sec:sami}

We selected a sample of 260 galaxies from the SAMI Galaxy Survey internal data release version~0.9. The Sydney-AAO Multi-object Integral field spectrograph \citep[SAMI;][]{CroomEtAl2012} is a front-end fibre feed system for the AAOmega spectrograph \citep{SharpEtAl2006}, consisting of 13~bundles of 61~fibres each \citep[`hexabundles';][]{BlandHawthornEtAl2011,BryantEtAl2014} that can be deployed over a 1~degree diameter field of view. SAMI therefore enables simultaneous spatially-resolved spectroscopy of twelve galaxies and one calibration star with a 15''~diameter field-of-view on each object.  The AAOmega spectrograph can be configured to provide different resolutions and wavelength ranges; the SAMI Galaxy survey employs the 570V grating to obtain a resolution of $\mathrm{R}=1730$ ($74\,\km\,\s^{-1}$) at \mbox{$3700$--$5700\,\mathrm{\AA}$} and the 1000R grating to obtain $\mathrm{R}=4500$ ($29\,\km\,\s^{-1}$) at \mbox{$6250$--$7350\,\mathrm{\AA}$}. SAMI datacubes are reduced and re-gridded to a spatial scale of $0.5''\times0.5''$ \citep{SharpEtAl2015} and the spatial resolution is about $2''$ (Green et al., in prep).

The SAMI Galaxy Survey plans to include more than 3000 galaxies at redshift $z<0.1$ covering a wide range of stellar masses and environments.  The sample is drawn from GAMA \citep{DriverEtAl2011} with additional entries from eight nearby clusters to cover denser environments \citep[][Owers et al., in prep]{BryantEtAl2015}. Reduced datacubes and a variety of emission line based higher-level data products are included in the first public data release \citep[][Green et al., in prep]{AllenEtAl2015}.

The emission lines of SAMI galaxies have been analysed using the spectral fitting pipeline LZIFU \citep{HoEtAl2016} to extract emission line fluxes and kinematics for each spectrum. The spectrum associated with each spectral/spatial pixel (`spaxel') is first fit with a stellar template using the `penalized pixel-fitting' (pPXF) routine \citep{CappellariEmsellem2004,Cappellari2017} before fitting up to three Gaussian line profiles to each of eleven strong emission lines simultaneously. For this paper, we choose to use the single Gaussian fits, and make use of the emission line flux maps, gas velocity maps, and gas velocity dispersion maps below.

Also available in the SAMI Galaxy Survey database are maps of SFR and $\sigsfr$ (in units of $\sigsfrunit$). These maps are made using extinction-corrected H$\alpha$ fluxes converted to SFRs following the relation derived in \citet{KennicuttEtAl1994}. The SFR maps are fully described in Medling et al. (in prep).

\subsection{Our subsample} \label{sec:oursubsample}

\begin{figure*}
\centerline{\includegraphics[width=1.0\linewidth]{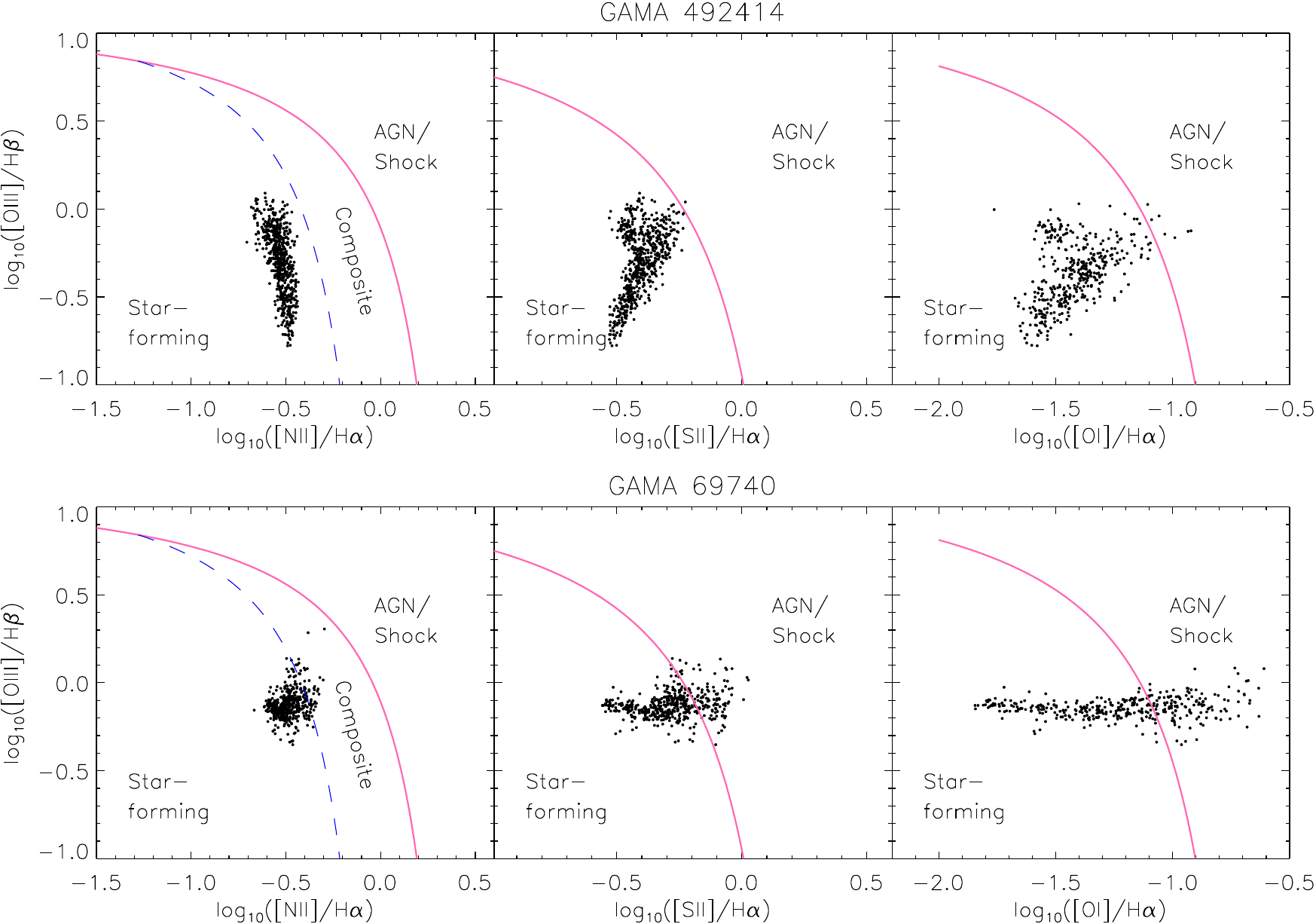}}
\caption{Emission line diagnostic diagrams \citep*[BPT/VO diagrams;][]{Baldwin:1981aa,VeilleuxOsterbrock1987} for galaxy GAMA~492414 (\emph{top panels}), which is classified as a Star-forming galaxy, and galaxy GAMA~69740 (\emph{bottom panels}), which is classified as a {\cas} galaxy. Each point in these diagrams corresponds to a spaxel from the spatial map of the data cube. \emph{Left-hand panels:} [O\texttt{III}]/H$\beta$ versus [N\texttt{II]}/H$\alpha$ diagnostic diagram, in which star-forming spaxels lie below the dashed line \citep{KauffmannEtAl2003,K06}. Points lying above the solid line are AGN-dominated \citep{KewleyEtAl2001}, whilst those lying in between the two regions experience significant contributions from both star formation and AGN activity, and are thus classified as being Composite \citep{K06}. \emph{Middle and right-hand panels:} [O\texttt{III}]/H$\beta$ versus [S\texttt{II}]/H$\alpha$ and [O\texttt{III}]/H$\beta$ versus [O\texttt{I}]/H$\alpha$ diagnostic diagrams, respectively. In both cases, points falling below the solid line are classified as star-forming, whereas those lying above are AGN or shock-dominated.}
\label{fig:bpt}
\end{figure*} 

From the pool of SAMI galaxies, we select a sub-sample of galaxies according to the criteria described below. We only consider spaxels with a sufficiently high signal-to-noise (S/N) ratio. The S/N was defined to be the ratio of the total emission line flux to the statistical one-sigma error in the line flux. This error was inferred using the Levenberg-Marquardt technique of chi-squared minimisation \citep{HoEtAl2016}. In the following, we list the selection criteria:

\begin{enumerate}

\item Source Extractor (SExtractor) ellipticity values are available. These values were obtained from the GAMA database \citep{Driver:2009aa,Baldry:2010aa,Robotham:2010aa,DriverEtAl2011,Hopkins:2013aa,Baldry:2014aa}. We require the ellipticity for each galaxy to estimate the physical volume of gas within each spaxel (explained in detail in Sec.~\ref{sec:tff} below).

\item The S/N ratio must be $\geq5$ in the H$\alpha$, H$\beta$, [N\texttt{II}], [S\texttt{II}], [O\texttt{I}] and [O\texttt{III}] emission lines. This allows reliable classification of the emission mechanism. However, in order to measure velocity dispersions down to about $12\,\km\,\s^{-1}$, we require and impose an S/N ratio of $\geq34$ in the measured velocity dispersion (explained in detail in Sec.~\ref{subsec:veldisp} below). We also require that beam-smearing (see Sec.~\ref{subsec:veldisp}) did not have a significant effect on the measured velocity dispersion.

\item After removing spaxels that have low S/N and/or are affected by beam-smearing, the galaxy must have more than ten star-forming spaxels remaining. The star-forming spaxels were filtered using the optical classification criteria given in \citet{K06}, an example of which is shown in Fig.~\ref{fig:bpt}. This classification scheme uses optical emission line ratios \citep*[BPT/VO diagrams;][]{Baldwin:1981aa,VeilleuxOsterbrock1987}, in order to distinguish between star-forming galaxies and galaxies that are dominated by an active galactic nucleus (AGN) or by shocks. The H$\alpha$-to-SFR conversion factor used in this work is only valid for star-forming regions, because AGN/shock-dominated spaxels are contaminated with emission from AGN/shock regions \citep{Kewley:2002aa,Kewley:2003aa,K06,Rich:2010aa,Rich:2011aa,Rich:2012aa}.

\end{enumerate}

Emission line fluxes of each spaxel were corrected for extinction using the Balmer decrement and the \citet{Cardelli:1989aa} reddening curve. Standard extinction for the diffuse ISM was assumed, with an $R_v$ value of 3.1 being utilised throughout the analysis \citep{Cardelli:1989aa,Calzetti:2000aa}.

Each galaxy was classified as either a `Star-forming' or `{\cas}' galaxy. To be classified as Star-forming, the galaxy had to have at least $90\%$ of all valid spaxels lying below and to the left-hand side of the \citet{KauffmannEtAl2003} classification line in the [O\texttt{III}]/H$\beta$ versus [N\texttt{II}]/H$\alpha$ diagram, and below and to the left-hand side of the \citet{KewleyEtAl2001} line in the [S\texttt{II}]/H$\alpha$ and [O\texttt{I}]/H$\alpha$ diagrams, as described in \citet{K06} (see Fig.~\ref{fig:bpt}). A galaxy was classified as {\cas}, if at least $10\%$ of all valid spaxels lie above the \citet{KauffmannEtAl2003} classification line on the [O\texttt{III}]/H$\beta$ versus [N\texttt{II}]/H$\alpha$ diagram and above the \citet{KewleyEtAl2001} classification line on the [S\texttt{II}]/H$\alpha$ and [O\texttt{I}]/H$\alpha$ diagnostic diagrams. Thus, {\cas} galaxies may include Composite, AGN, or Shock \citep{Kewley:2013aa} galaxies according to the classification in \citet{K06}. These classifications resulted in a sample of 219~Star-forming and 41~{\cas} classified galaxies.


\section{Estimating the molecular gas surface density ($\siggas$)} \label{sec:theory}

\begin{figure*}
\centerline{\includegraphics[width=1.0\linewidth]{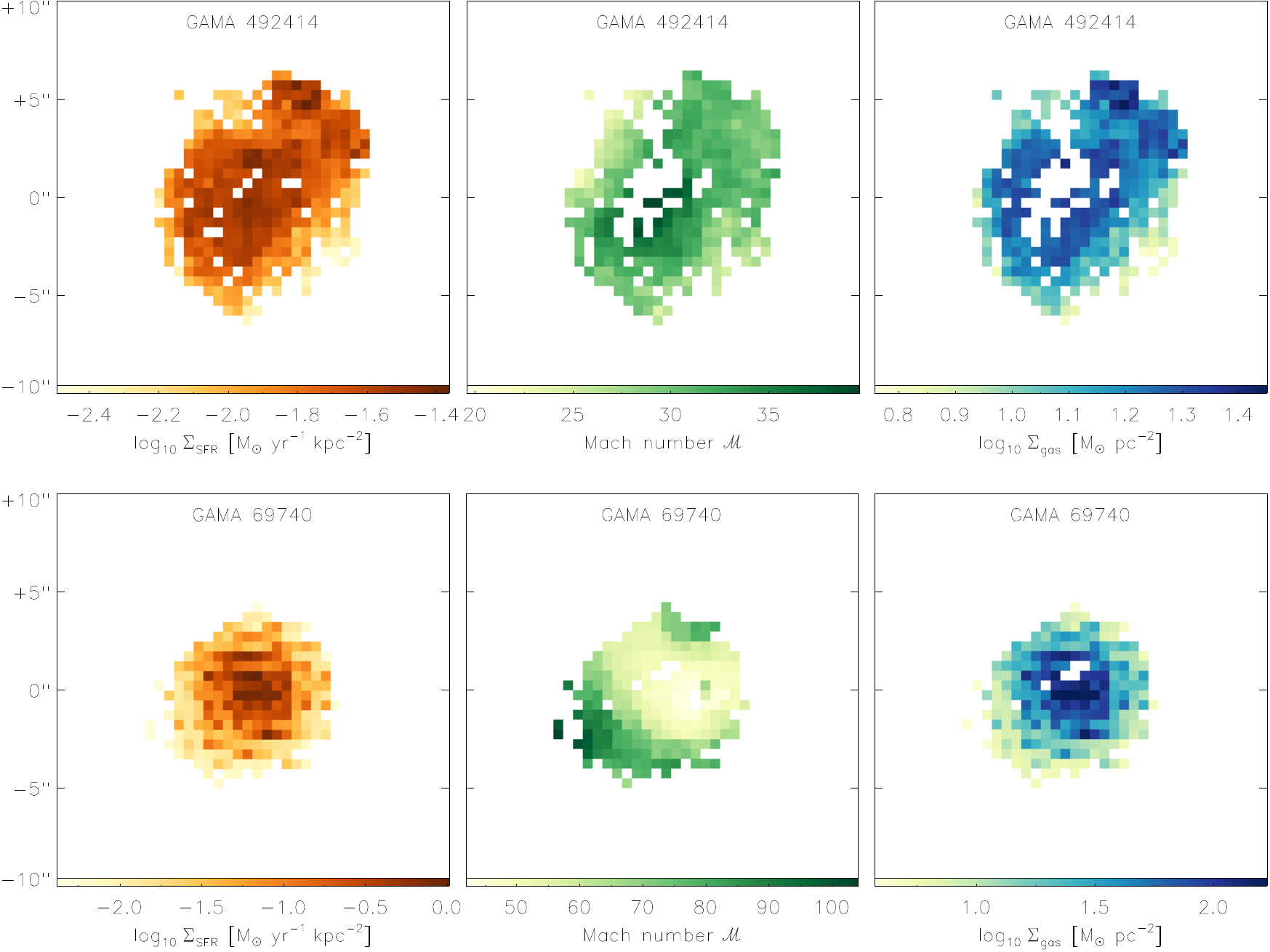}}
\caption{Maps depicting the inputs and outputs of our method of estimating $\siggas$ for example galaxy GAMA~492414, a Star-forming classified galaxy (\emph{top panels}) and GAMA~69740, a {\cas} classified galaxy (\emph{bottom panels}). In both cases, only spaxels that were classified as dominated by star formation are analysed, and AGN or shock-dominated spaxels are excluded from the $\siggas$ reconstruction. \emph{Left-hand panels:} $\sigsfr$ maps obtained from the SAMI H$\alpha$ flux (Sec.~\ref{sec:sami}). This map is the first input to obtaining our $\siggas$ prediction. \emph{Middle panels:} Mach number ($\mathcal{M}$) map obtained from the SAMI H$\alpha$ velocity dispersion under the assumptions and procedures outlined in Sec.~\ref{sec:mach}. The Mach number maps are our second input. We note that the Mach number maps often have spaxels missing towards the centre of the galaxies---this is because of the relatively aggressive S/N cuts of 34 on the velocity dispersion and because of our conservative beam-smearing cut of all spaxels with $\sigma_v < 2\,v_\mathrm{grad}$ (see Sec.~\ref{sec:mach} for details). \emph{Right-hand panels:} prediction for the distribution of molecular gas column density ($\siggas$), which is our final product.} 
\label{fig:maps}
\end{figure*}

Here we exploit the spatially resolved SAMI H$\alpha$ flux and $\sigsfr$ maps in combination with the H$\alpha$ velocity dispersion maps to derive predictions of $\siggas$ across each galaxy in our sample. Examples of $\sigsfr$ maps are shown in the left-hand panels of Fig.~\ref{fig:maps} for the Star-forming and {\cas} classified galaxies from Fig.~\ref{fig:bpt}. We further derive spaxel-averaged values of the physical parameters for each galaxy in our subsample.

\subsection{Deriving turbulent Mach number maps ($\mathcal{M}$)} \label{sec:mach}

The SFK15 model relies on the availability of the sonic Mach number $\mathcal{M}$, the ratio between the gas velocity dispersion and the local speed of sound. This was prompted by the findings of \citet{Federrath2013sflaw}, showing that the observed scatter within the K98 and KDM12 relations may be primarily attributed to the physical variations in $\mathcal{M}$. As direct measurements of $\mathcal{M}$ are unavailable for the SAMI sample, for every pixel we estimate a value using the method described in the following sections. The result of these procedures is shown in the middle panels of Fig.~\ref{fig:maps}, for the two example galaxies, GAMA~492414 (top) and GAMA~69740 (bottom).

\subsubsection{Estimating the sound speed}
Molecular clouds in Galactic spiral arms exhibit a gas temperature range of \mbox{$T\sim10$--50$\,\mathrm{K}$}, while those in the Galactic centre can have temperatures up to $100\,\mathrm{K}$ \citep{GinsburgEtAl2016}. All $\mathrm{H_2}$ gas should lie within this temperature range, otherwise it will cease to be molecular under typical conditions in the ISM \citep{Ferriere:2001aa}. The local sound speed ($\cs$) of the gas is given by
\begin{equation} \label{eq:cs}
\cs=\left[k_\mathrm{B}T/\left(\mu_\mathrm{p}m_\mathrm{H}\right)\right]^{1/2},
\end{equation}
with the Boltzmann constant $k_\mathrm{B}$, the mass of a hydrogen atom $m_\mathrm{H}$, and the mean particle weight $\mu_\mathrm{p}$. The latter is $\mu_\mathrm{p}=2.3$ for molecular gas and $\mu_\mathrm{p}=0.6$ for ionised gas, assuming standard cosmic abundances \citep{KauffmannEtAl2008}. Therefore temperatures of $T=10\,\mathrm{K}$ and $T=100\,\mathrm{K}$ correspond to molecular sound speeds of $\cs=0.2\,\km\,\s^{-1}$ and $0.6\,\km\,\s^{-1}$, respectively. We hence estimate the Mach number of the gas in each spaxel by dividing the velocity dispersion by the molecular sound speed of $\cs=0.4\pm0.2\,\km\,\s^{-1}$, which is appropriate for the dense, cold star-forming phase of the ISM in the temperature range \mbox{$T\sim10$--$100\,\mathrm{K}$}. 

\subsubsection{Turbulent velocity dispersion} \label{subsec:veldisp}
In order to apply our star formation relation, Eq.~(\ref{eq:newsflaw}), we need an estimate of the turbulent velocity dispersion of the molecular gas in order to construct the turbulent Mach number. Here we use the H$\alpha$ velocity dispersion to approximate the velocity dispersion of the cold gas. The H$\alpha$ velocity dispersion is similar (to within a factor of \mbox{$2$--$3$}) to the molecular gas velocity dispersion, because of the coupling of turbulent gas flows between the hot, warm and cold phases of the ISM. For instance, it has been found that for M33, the second-most luminous spiral galaxy in our local group, the atomic H\texttt{I} dispersions are a fair estimator of the CO dispersions \citep{Druard:2014aa}. In M33, H$\alpha$ velocities have been found to trace H\texttt{I} velocities reasonably well \citep{KamEtAl2015}. However, the H$\alpha$ velocity dispersion is expected to be somewhat higher than the H$_2$ velocity dispersion, because the ionised emission comes from H\texttt{II} regions close to massive stars, which directly contribute to driving turbulence. We thus expect the velocity dispersion in the direct vicinity of massive stars to be overestimated. In order to take this effect into account, we crudely approximate the H$_2$ velocity dispersion with half the H$\alpha$ velocity dispersion, $\sigma_v=\sigma_v(\mathrm{H}\alpha)/2$. While this provides only a rough estimate of the molecular velocity dispersion (trustworthy only to within a factor of 2--3), we show below that the uncertainties that this introduces into our $\siggas$ estimate are only of the order of $50\%$. This is because of the relatively weak dependence of $\siggas$ on $\mach$, as we will derive in Sec.~\ref{sec:uncertainties} below. To demonstrate this, we investigate a case below, where we assume that the molecular gas velocity dispersion is equal to the H$\alpha$ velocity dispersion, $\sigma_v=\sigma_v(\mathrm{H}\alpha)$, which yields $<30\%$ lower $\siggas$. Thus, even though our velocity dispersion estimate is uncertain by factors of \mbox{$\sim2$--$3$}, the final uncertainty in $\siggas$ is $\lesssim50\%$.

\paragraph{S/N requirements:}

The SAMI/AAOmega spectrograph setup has an instrumental velocity resolution of $\sigma_\mathrm{instr}=29\,\km\,\s^{-1}$ at the wavelength of H$\alpha$ (see Sec.~\ref{sec:sami}). Velocity dispersions below this resolution limit can still be reliably measured if the S/N in the observed (instrument-convolved) velocity dispersion is sufficiently high. In the following, we estimate the required S/N in order to reconstruct intrinsic velocity dispersions down to $\sigma_\mathrm{true}=12\,\km\,\s^{-1}$. We choose this cutoff of $12\,\km\,\s^{-1}$, because it is the sound speed of the ionised gas, Eq.~(\ref{eq:cs}) with $T=10^4\,\mathrm{K}$ and $\mu_\mathrm{p}=0.6$, and thus represents a physical lower limit for $\sigma$.

The intrinsic (true) velocity dispersion ($\sigma_\mathrm{true}$) can be obtained by subtracting the instrumental velocity resolution ($\sigma_\mathrm{instr}$) from the observed (instrument-convolved) velocity dispersion ($\sigma_\mathrm{obs}$) in quadrature, with
\begin{align}
\sigma_\mathrm{true}^2 = \sigma_\mathrm{obs}^2 - \sigma_\mathrm{instr}^2. \label{eq:S_N_cuts}
\end{align}
The same relation holds for the uncertainties (noise) in the velocity dispersion,
\begin{align}
d(\sigma_\mathrm{true}^2) & = d(\sigma_\mathrm{obs}^2) - d(\sigma_\mathrm{instr}^2), \nonumber \\
2\sigma_\mathrm{true}d(\sigma_\mathrm{true}) & = 2\sigma_\mathrm{obs}d(\sigma_\mathrm{obs}) - 2\sigma_\mathrm{instr}d(\sigma_\mathrm{instr}).
\end{align}
Assuming that the instrumental velocity resolution is fixed, we can use $d(\sigma_{\mathrm{instr}})=0$ and simplify the last equation to 
\begin{align}
d(\sigma_{\mathrm{true}}) &= \frac{\sigma_{\mathrm{obs}}}{\sigma_{\mathrm{true}}}  d(\sigma_{\mathrm{obs}}).
\end{align}
Dividing both sides by $\sigma_\mathrm{true}$ and substituting Eq.~(\ref{eq:S_N_cuts}) yields
\begin{align}
\frac{\sigma_\mathrm{obs}}{d(\sigma_\mathrm{obs})} = \frac{\sigma_\mathrm{true}}{d(\sigma_\mathrm{true})}\frac{\sigma_\mathrm{obs}^2}{\sigma_\mathrm{true}^2}
= \frac{\sigma_\mathrm{true}}{d(\sigma_\mathrm{true})} \left(1+\frac{\sigma_\mathrm{instr}^2}{\sigma_\mathrm{true}^2}\right).
\end{align}
Since $\mathrm{(S/N)_{obs}}\!\equiv\!\sigma_\mathrm{obs}/d(\sigma_\mathrm{obs})$ and $\mathrm{(S/N)_{true}}\!\equiv\!\sigma_\mathrm{true}/d(\sigma_\mathrm{true})$ are the observed (instrument-convolved) and intrinsic S/N ratios, respectively, we can estimate the required $\mathrm{(S/N)_{obs}}$ for the target intrinsic $\mathrm{(S/N)_{true}}=5$ and the target intrinsic velocity dispersion that we want to resolve, $\sigma_\mathrm{true}=12\,\km\,\s^{-1}$, by evaluating
\begin{align} \label{eq:sncutobs}
\mathrm{(S/N)_{obs}} & \geq \mathrm{(S/N)_{true}} \left(1+\frac{\sigma_\mathrm{instr}^2}{\sigma_\mathrm{true}^2}\right) \nonumber\\
                               & \geq 5 \left[1+\left(\frac{29\,\km\,\s^{-1}}{12\,\km\,\s^{-1}}\right)^2\right] = 34.
\end{align}
Thus, for spaxels with observed (instrument-convolved) velocity dispersion S/N ratios greater or equal to 34, we can reliably reconstruct the intrinsic (instrument-corrected) velocity dispersion down to $12\,\km\,\s^{-1}$, with an intrinsic S/N ratio of at least 5. We note that the SAMI database provides the instrument-subtracted velocity dispersion $\sigma_\mathrm{subtracted}$ (`\texttt{VDISP}') and its error $d(\sigma_\mathrm{subtracted})$ (`\texttt{VDISP\_ERR}') based on the LZIFU fits \citep{HoEtAl2016}. Thus, in order to apply the S/N cut of 34 derived in Eq.~(\ref{eq:sncutobs}), we first reconstruct $\sigma_\mathrm{obs}=(\sigma_\mathrm{subtracted}^2+\sigma_\mathrm{instr}^2)^{1/2}$ and its error $d(\sigma_{\mathrm{obs}})=d(\sigma_{\mathrm{subtracted}})\sigma_\mathrm{subtracted}/\sigma_\mathrm{obs}$, using error propagation. This criterion is functionally equivalent to setting a S/N cut on the instrument-subtracted velocity dispersion,
\begin{equation}
\mathrm{(S/N)_{subtracted}} = \frac{34}{1+\sigma_\mathrm{instr}^2/\sigma_\mathrm{subtracted}^2}.
\end{equation}

After applying our S/N cuts of 34 to the observed (instrument-convolved) velocity dispersion, any spaxels with velocity dispersions less than $12\,\km\,\s^{-1}$ are disregarded. We note that this final cut only removes 1\% of the spaxels with $\mathrm{(S/N)_{obs}}\geq34$.

\paragraph{Beam smearing:}

We also have to account for `beam smearing', a phenomenon that occurs because of the limitation in spatial resolution of the instrument. Beam smearing occurs for a physical velocity field that changes on spatial scales smaller than the spatial resolution of the observation. If there is a steep velocity gradient across neighbouring pixels, such as near the centre of a galaxy, beam smearing leads to an artificial increase in the measured velocity dispersion at such spatial locations. To account for beam smearing, we follow the method in \citet{Varidel:2016aa} and estimate the local velocity gradient $v_\mathrm{grad}$ for a given spaxel with coordinate indices $(i,j)$ as the magnitude of the vector sum of the difference in the velocities in the adjacent pixels,
\begin{align}
& v_\mathrm{grad}(i,j) = \\ \nonumber
& \quad\sqrt{[v(i+1,j)-v(i-1,j)]^2 + [v(i,j+1)-v(i,j-1)]^2}. 
\end{align}
Note that the differencing to compute $v_\mathrm{grad}$ occurs over a linear scale of three SAMI pixels along $i$ and $j$ and thus covers roughly the spatial resolution of the seeing-limited SAMI observations with $\mathrm{FWHM}\sim2''$ (see Sec.~\ref{sec:obs}). If a pixel has a neighbour that is undefined (e.g., because of low S/N), the gradient in that direction is not taken into account. As our standard criterion to account for beam smearing, we cut any pixels in which the velocity dispersion is less than twice that of the velocity gradient ($\sigma_v < 2\,v_\mathrm{grad}$) and disregard such pixels in further analyses, leaving only spaxels that are largely unaffected by beam smearing.

In addition to our fiducial beam-smearing criterion ($\sigma_v < 2\,v_\mathrm{grad}$), we test a case with a relaxed beam-smearing cut of $\sigma_v < v_\mathrm{grad}$, and find nearly identical results (see Tab.~\ref{tab:sum} below). We note that our standard beam-smearing cut with $\sigma_v < 2\,v_\mathrm{grad}$ tends to remove spaxels near the centre of some of the galaxies (see e.g., Fig.~\ref{fig:maps}). However, using the relaxed beam-smearing cut with $\sigma_v < v_\mathrm{grad}$ yields global (galaxy-averaged) Mach numbers and global $\siggas$ estimates that agree to within 4\% with our standard beam-smearing cut (see Tab.~\ref{tab:sum}), demonstrating that our results are largely unaffected by beam smearing.

\paragraph{Turbulent velocity dispersion versus systematic motions:}

Beam-smearing is the result of un-resolved velocity gradients in the plane-of-the-sky. However, systematic velocity gradients (such as resulting from rotation or large-scale shear) along the line-of-sight (LOS) also increase the velocity dispersion (even for arbitrarily high spatial resolution) by LOS-blending. These large-scale systematic motions do not represent turbulent gas flows \citep[see e.g., the recent study of turbulent motions in the Galactic-centre cloud `Brick', which is subject to large-scale shear][]{FederrathEtAl2016}. As we have not subtracted or accounted for these factors, our values of the turbulent velocity dispersion may be overestimated.

In summary, we emphasise that the turbulent velocity dispersion has large uncertainties and is only accurate to within a factor of $2$--$3$. However, the uncertainties that this introduces into our final product ($\siggas$) are $\lesssim50\%$, because of the relatively weak dependence of $\siggas$ on $\mach$ (derived in detail in Sec.~\ref{sec:uncertainties} below).

\subsection{Deriving $\gastmff$ and $\gastff$} \label{sec:gastff}
To find the MGCR $\gastmff$, we divide $\sigsfr$ (left-hand panels of Fig.~\ref{fig:maps}) by the SFR efficiency of 0.45$\%$ found in SFK15. That is, we invert Eq.~(\ref{eq:newsflaw}),
\begin{align} \label{eq:gastmff}
\gastmff\,\left[\sigsfrunit\right] = \frac{\sigsfr}{0.0045} .
\end{align}
In order to find the ratio between the gas column density and the freefall time at the average gas density, $\gastff$, we take the Mach number calculated in Sec.~\ref{sec:mach} and convert $\gastmff$ to $\gastff$,
\begin{align} \label{eq:gastff}
\gastff\,\left[\sigsfrunit\right] = \frac{\gastmff}{\left(1 + b^2 \mathcal{M}^2 \frac{\beta}{\beta+1}\right)^{3/8}}.
\end{align}
In the following, we will assume a fixed turbulence driving parameter $b=0.4$, representing a natural mixture \citep{FederrathKlessenSchmidt2008}, and assume an absence of magnetic fields such that $\beta\to\infty$. Although both of these are strong assumptions, we emphasise that, in the absence of constraints on $b$ or $\beta$ in these galaxies, we have to assume fixed, typical values for them and allow that these assumptions contribute to the uncertainties of the $\siggas$ estimation. However, if these parameters will be measured in the future, they can be used in Eqs.~(\ref{eq:newsflaw}) and~(\ref{eq:gastff}) to obtain a more accurate prediction of $\siggas$. For simplicity, here we fix $b$ and $\beta$, and only consider the remaining dependence on $\mach$.

\subsection{Estimating the gas density ($\rho$) and local freefall time ($\tff$)} \label{sec:tff}

Now that we have $\gastff\equiv\siggas/\tff$ from Eq.~(\ref{eq:gastff}), we need an estimate of the average freefall time $\tff=\sqrt{3\pi/(32G\rho)}$ to obtain $\siggas$ from $\gastff$. Thus, we need an estimate of the local gas density $\rho$, which requires some geometrical considerations and assumptions similar to the ones outlined in KDM12.

\begin{figure}
\centerline{\includegraphics[width=0.9\linewidth]{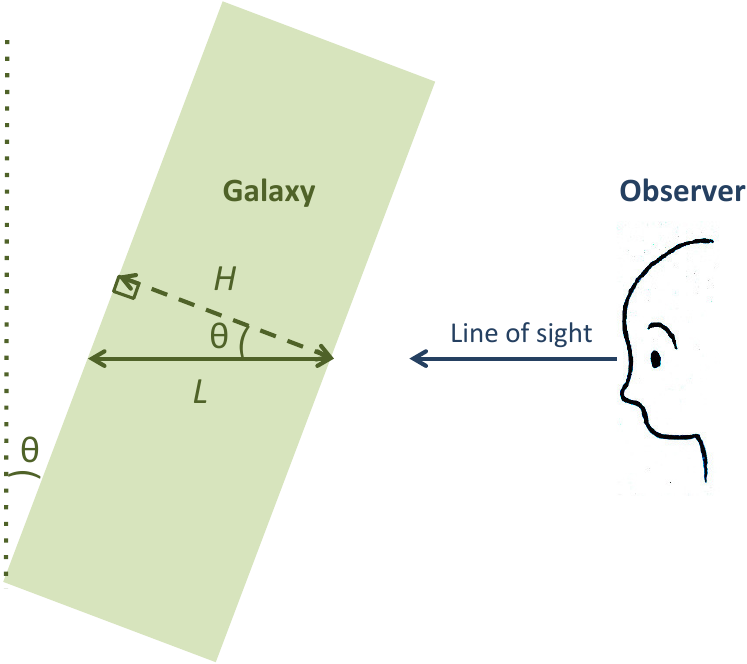}}
\caption{Schematic diagram of our model for each galaxy's geometry from Eq.~(\ref{eq:L}). In this diagram the green rectangle represents a section of a galaxy with scale height $H$, which is inclined an angle $\theta$ away from the vertical. $L$ is the column length used to estimate the local gas density and freefall time.}
\label{fig:geometry}
\end{figure}

First, we make the assumption that the galaxy has a uniform gas disc geometry with a scale height of \mbox{$H=(100\pm50)\,\pc$} \citep{Glazebrook:2013aa, van-der-Kruit:2011aa}. However, depending on the viewing angle with respect to the orientation of the galactic disc in the plane of the sky, the LOS length through the gas may be greater than the scale height (see Fig.~\ref{fig:geometry}). This angle can be estimated from the observed ellipticity of the galaxy. To correct for the viewing angle, we obtain SExtractor ellipticity values, $\varepsilon$, for each galaxy from the GAMA database, from which we obtain the inclination angle, $\theta$, of the galaxy:
\begin{align}
\theta = \arccos{(1 - \varepsilon)}.
\end{align}
The column length, $L$, can then be inferred by dividing the scale height, $H$, by the cosine of the inclination angle, as pictured in Fig.~\ref{fig:geometry},
\begin{align}
L \,\left[\pc\right] = \frac{H}{\cos(\theta)} = \frac{100}{\cos(\theta)} = \frac{100}{1 - \varepsilon}. \label{eq:L}
\end{align}
We caution that the assumed cylindrical geometry is a drastic simplification, as there has been much evidence to suggest that the scale height of a galaxy follows a relation dependent upon its radius from the galactic centre \citep{Toomre:1964aa,van-der-Kruit:1981aa,van-der-Kruit:1982aa,de-Grijs:1996aa,de-Grijs:1997aa}. This may cause our predicted gas maps to underestimate the gas density towards the centre and overestimate the gas density towards the outskirts of the galaxy. However, the general shape of the predicted distribution of gas and especially the galaxy-averaged gas surface density should not be affected significantly by this geometrical simplification. A refinement in the geometry is relatively straightforward to implement, if one requires more accurate maps. We estimate that the relative uncertainties in $L$ may be up to $100\%$. However, our final result ($\siggas$), does not depend significantly on $L$ (see detailed discussion in Sec.~\ref{sec:uncertainties}).

Given the column length $L$, we can write the gas density
\begin{align}
\rho = \frac{\siggas}{L}.
\label{gas_rho_L}
\end{align} 
Since we do not have $\siggas$ because it is our final product, we now substitute a rearrangement of the definition of $\gastff$,
\begin{align}
\siggas = \gastff \times t_{\mathrm{ff}}, \label{eq:gas_single_tff}
\end{align}
as well as the definition of the freefall time in terms of $\rho$,
\begin{align}
t_{\mathrm{ff}}(\rho)=\sqrt{\frac{3\pi}{32G\rho}}, \label{eq:tff}
\end{align}
where $G$ is the gravitational constant. We combine the three previous equations and solve for the gas density,
\begin{align}
\rho &=\frac{\gastff}{L} \sqrt{\frac{3\pi}{32G\rho}}\\
\Rightarrow \rho &= \left(\sqrt{\frac{3\pi}{32G}}\times\frac{\gastff}{L}\right)^{2/3}. \label{eq:rho}
\end{align}
We substitute $\rho$ back into Eq.~(\ref{eq:tff}) to obtain the freefall time $t_{\mathrm{ff}}$ for the average gas density $\rho$.

\subsection{Deriving our final product, $\siggas$}

Finally, we obtain our prediction for $\siggas$ either by multiplying the freefall time from Sec.~\ref{sec:tff} by $\gastff$ calculated in Sec.~\ref{sec:gastff}, i.e., using Eq.~(\ref{eq:gas_single_tff}), or by multiplying the volume density $\rho$ from Eq.~(\ref{eq:rho}) by the column length $L$ from Eq.~(\ref{eq:L}). In terms of the principle observables, $\sigsfr$ and $\mathcal{M}=\sigma_v/\cs$, as well as our assumptions for the parameters $L=H/(1-\varepsilon)$, $b$ and $\beta$, this corresponds to the final expression for $\siggas$ given by
\begin{align} \label{eq:gas}
\siggas = \left(\frac{3\pi L}{32G}\right)^{\frac{1}{3}} \left[\frac{\sigsfr}{0.0045\left(1 + b^2 \mathcal{M}^2 \frac{\beta}{1+\beta}\right)^{3/8}}\right]^{2/3}.
\end{align}
Two examples of the spatially resolved maps of estimated $\siggas$ based on the new method provided by Eq.~(\ref{eq:gas}) are shown in the right-hand panels of Fig.~\ref{fig:maps}.

\subsection{Uncertainties in the $\siggas$ reconstruction} \label{sec:uncertainties}

Here we estimate the uncertainties in our $\siggas$ prediction based on Eq.~(\ref{eq:gas}). We derive the uncertainties by error propagation of all variables in Eq.~(\ref{eq:gas}). First, we note that the dependence of $\siggas$ on $L$ is weak ($\siggas\propto L^{1/3}$) and the dependence on $\mach$ is also relatively weak ($\siggas\propto\mach^{-1/2}$), which means that the uncertainties in $L$ and $\mach$ enter the final uncertainty in $\siggas$ with a weight of 1/3 and 1/2, respectively. The strongest dependence of $\siggas$ is on the SFR, i.e., $\siggas\propto\sigsfr^{2/3}$, so the uncertainties in $\sigsfr$ are weighted by 2/3, and we thus expect these to dominate the final uncertainties. Rigorously, the relative uncertainty $\err(\siggas)/\siggas$ from Eq.~(\ref{eq:gas}) is given by
\begin{align} \label{eq:gas_uncertainty}
& \frac{\err(\siggas)}{\siggas} =  \\ \nonumber
& \quad \left[ \left(\frac{1}{3}\frac{\err(L)}{L}\right)^2 + \left(\frac{1}{2}\frac{\err(\mach)}{\mach}\right)^2 + \left(\frac{2}{3}\frac{\err(\sigsfr)}{\sigsfr}\right)^2 \right]^{1/2},
\end{align}
where we approximated the denominator $(1+b^2\mach^2)$ in Eq.~(\ref{eq:gas}) as $b^2\mach^2$ for the uncertainty propagation (recall that we also assumed $\beta\!\to\!\infty$), because $b^2\mach^2\gg1$, based on our velocity dispersion cut and sound speed (see Sec.~\ref{subsec:veldisp}). With typical relative uncertainties of 70\% in $L$, $100\%$ in $\mach$ (see Sec.~\ref{sec:mach}) and $20\%$ in $\sigsfr$ (based on our S/N cuts of 5 on the H$\alpha$ flux; see Sec.~\ref{sec:oursubsample}), we find a relative uncertainty of $\err(\siggas)/\siggas=57\%$, which is dominated by the uncertainty in $\sigsfr$. Even if the uncertainties in both $L$ and $\mach$ were $100\%$ and $150\%$ respectively, we would still be able to estimate $\siggas$ with an uncertainty of $83\%$. In summary, despite the large uncertainties in $\mach$ and $L$ (see Sec.~\ref{sec:mach} and~\ref{sec:tff}), our final uncertainties in $\siggas$ are less than a factor of 2.


\section{Results} \label{sec:results}

\subsection{Gas surface density estimates} 

\begin{figure*}
\centerline{\includegraphics[width=1.0\linewidth]{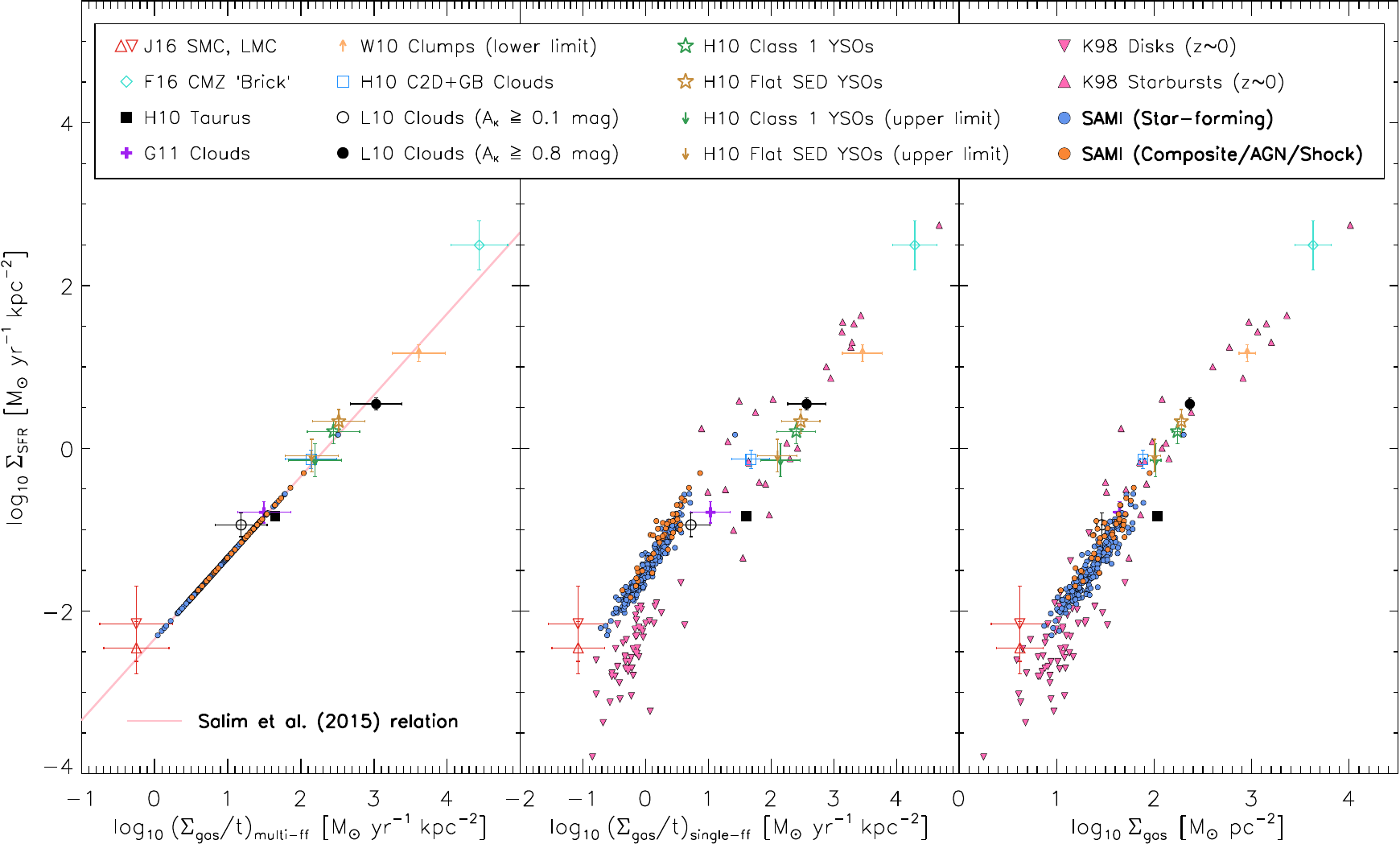}}
\caption{\emph{Left-hand panel:} $\sigsfr$ versus $\gastmff$, i.e., the star formation relation derived in SFK15 (Eq.~\ref{eq:newsflaw}; solid line). The data points shown are the log-averaged observational data used to derive the SFK15 relation: \citet{HeidermanEtAl2010} (H10), \citet{GutermuthEtAl2011} (G11), \citet{WuEtAl2010} (W10), and \citet{LadaLombardiAlves2010} (L10). Also shown are updated and new observational data based on recent works for the SMC and LMC \citep{JamesonEtAl2016} (J16), and for the CMZ cloud `Brick' \citep{FederrathEtAl2016,BarnesEtAl2017} (F16). Error bars are the standard deviation of the mean of the individual cloud data, except for the CMZ cloud `Brick', where the error is taken straight from the measurement.
\emph{Middle panel:} same as left-hand panel, but showing $\sigsfr$ versus $\gastff$ (KDM12 relation). We additionally include the individual K98 disc and starburst galaxies tabulated in KDM12 \citep[taking into account the corrections by][]{Federrath2013sflaw,KDM13}.
\emph{Right-hand panel:} same as middle panel, but showing $\sigsfr$ versus $\siggas$, where $\siggas$ for the SAMI galaxies (shown as filled circles in blue for Star-forming and orange for {\cas}) was estimated based on Eq.~(\ref{eq:gas}). Our estimates of $\siggas$ for SAMI lie in close proximity of the low-redshift K98 disc galaxies (filled down-pointing triangles).}
\label{fig:derivation}
\end{figure*}

Our main objective is to estimate $\siggas$ from $\sigsfr$ and the turbulence properties ($\mach$) in our SAMI galaxy sample. We do this by applying the new method introduced in the previous section (Sec.~\ref{sec:theory}), going step-by-step from $\sigsfr$ to $\siggas$.

Fig.~\ref{fig:derivation} shows each of the $\sigsfr$ parameterisations explored in SFK15, presented in the same order as the computations of our $\siggas$ derivations (Sec.~\ref{sec:theory}). The framework of the first panel assumes a direct correlation between $\sigsfr$ and $\gastmff$. That is, it assumes the star formation relation of Eq.~(\ref{eq:newsflaw}) to hold, thus by construction the SAMI data points in this framework lie along the same line. The data points from SFK15 which were used to obtain this relation are also shown. We note that in the SFK15 derivation of Eq.~(\ref{eq:newsflaw}), the K98 galaxies were omitted because they did not have $\gastmff$ values assigned to them due to their lack of $\mach$ measurements. They are thus similarly excluded in this panel.

Compared to the observational data published in SFK15, we updated and corrected some of the previous data, and added new observations in Figure~\ref{fig:derivation}. First, we replace the \citet{BolattoEtAl2011} data for the SMC by the most recent $200\,\pc$ resolution data provided in \citet{JamesonEtAl2016} (J16). We also add the Large Magellanic Cloud (LMC) data from \citet{JamesonEtAl2016} and assume that the SMC and LMC data have Mach numbers in the range \mbox{10--100}, i.e., we basically treat the Mach number as unconstrained, i.e., varying in a plausible range, but we currently do not have direct measurements of $\mach$ in the SMC or LMC.\footnote{The Mach number range of \mbox{16--200} assumed in SFK15 for the SMC was somewhat too high, because the $200\,\pc$ resolution data from \citet{BolattoEtAl2011} and \citet{JamesonEtAl2016} are more consistent with velocity dispersions that correspond to \mbox{$\mach\sim10$--$100$} for the SMC and LMC. However, without a direct measurement of the velocity dispersion and gas temperature, the Mach number remains rather unconstrained for the SMC and LMC.} Second, we replace the global CMZ data from \citet{YusefZadehEtAl2009} by the local CMZ cloud G0.253+0.016 `Brick' \citep{FederrathEtAl2016,BarnesEtAl2017} for which significantly more information is available. We take the values of $\siggas$, $\mach$, $b$, and $\beta$ measured in \citet{FederrathEtAl2016} (F16) and use the SFR per freefall time estimate of 2\% from \citet{BarnesEtAl2017} to obtain $\sigsfr$ for the `Brick'. The other cloud data are identical to those published in KDM12, F13 and SFK15, which were taken from \citet{HeidermanEtAl2010} (H10), \citet{GutermuthEtAl2011} (G11), \citet{WuEtAl2010} (W10), and \citet{LadaLombardiAlves2010} (H10). However, we corrected the error bar on the L10 clouds, which showed the standard deviation instead of the standard deviation of the mean in SFK15. We further propagated the uncertainties in $\sigsfr$, $\mach$ and $\tff$ between $\gastmff$, $\gastff$ and $\siggas$. Finally, we note that the observational data included in Fig.~\ref{fig:derivation} cover a wide range in spatial and spectral resolution (for details we refer the reader to the source publications of these data), which allowed us to test the universality of the SFK15 relation. In the future, when turbulence estimates become available for high-redshift data, those need to be included as well, to revisit the question of universality of the star formation relation derived in SFK15.

The second panel of Fig.~\ref{fig:derivation} depicts the KDM12 parameterisation, $\sigsfr$ versus $\gastff$. The derivation of this value for the SAMI galaxies required inputs from both the H$\alpha$ flux and velocity dispersion, with $\gastff$ computed from Eq.~(\ref{eq:gastff}). In addition to the observational data shown in the left-hand panel, we added the individual K98 disc and starburst galaxies from KDM12 \citep[with corrections based on][]{Federrath2013sflaw,KDM13}.

The third panel of Fig.~\ref{fig:derivation} shows the final product of our $\siggas$ predictions; the average gas column density estimate for each of the SAMI galaxies in our sample. These predictions span a range of \mbox{$\log_{10}\siggas\,[\siggasunit] \sim0.9$--$2.3$}. We note that the estimated $\siggas$ values for the SAMI galaxies are close to the $\siggas$ values of the K98 low-redshift disc galaxies. This is encouraging, because they are the most similar in type to our sample of SAMI galaxies. The offset in $\siggas$ and $\sigsfr$ by $\sim0.5\,\mathrm{dex}$ between the SAMI and K98 galaxies can be understood as a consequence of spatial resolution. In contrast to our spaxel-resolved analysis of the SAMI galaxies (with spatial resolution of $\sim2''$; see Sec.~\ref{sec:sami}), the K98 galaxies are unresolved, which reduces the inferred $\sigsfr$ \citep{FederrathKlessen2012,KruijssenLongmore2014,FisherEtAl2017}. The reason is that, although the total H$\alpha$ flux ($\propto\mathrm{SFR}$) remains similar even at lower resolutions, the area $\Delta A$ over which H$\alpha$ is emitted tends to be overestimated and hence the $\sigsfr$ tends to be underestimated ($\sigsfr=\mathrm{SFR}/\Delta A$) for the global K98 data. Similar holds for $\siggas$, because it depends on $\Delta A$ in the same way as $\sigsfr$, and indeed, we find that the resolved SAMI galaxies tend to lie at somewhat higher $\siggas$ compared to the unresolved K98 disc galaxy sample.

\subsection{Comparison between Star-forming and {\cas} galaxies}

\subsubsection{Gas surface density in Star-forming and {\cas} galaxies}

\begin{figure}
\centerline{\includegraphics[width=1.0\linewidth]{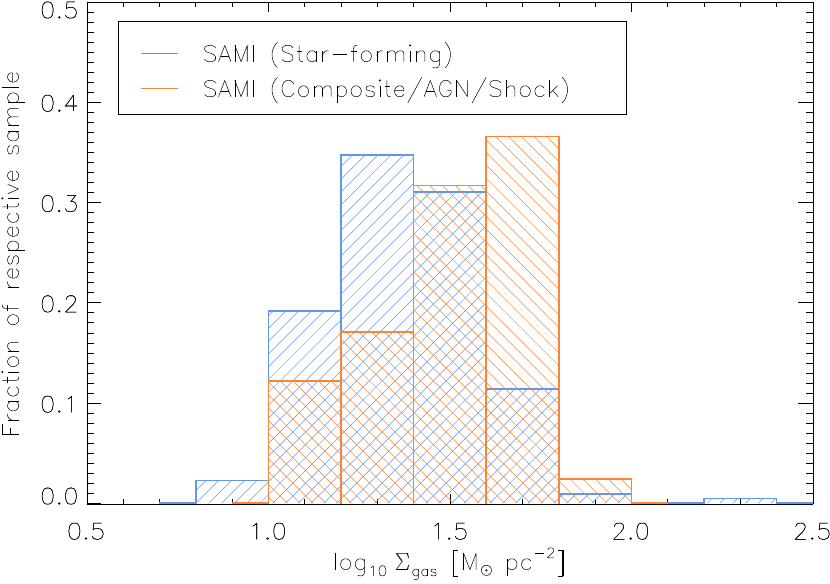}}
\caption{Normalised histograms showing the distributions of predicted $\siggas$ for SAMI galaxies classified as pure Star-forming galaxies (blue) and {\cas} galaxies (orange).}
\label{fig:gas_histograms}
\end{figure}

\begin{table*}
\caption{Average physical parameters for Star-forming and {\cas} classified SAMI galaxies.}
\def\arraystretch{1.0}
\setlength{\tabcolsep}{1.0pt}
\begin{tabular}{lcc}
\hline
Physical parameter \hspace{2.0cm} & Average (standard deviation) & \hspace{0.0cm}Average (standard deviation) \\
                                \hspace{2.0cm} & for Star-forming galaxies  & \hspace{0.0cm}for {\cas} galaxies \\
\hline
$\sigsfr\,[\sigsfrunit]$ & $0.054\,(0.11)$ & $0.11\,(0.09)$ \\
\hline
\multicolumn{3}{l}{\emph{Using the fiducial model: beam-smearing cut of $\sigma_v < 2\,v_\mathrm{grad}$, and estimate of the molecular velocity dispersion, $\sigma_v=\sigma_v(\mathrm{H}\alpha)/2$}} \vspace{0.05cm} \\
Mach number ($\mathcal{M}$) & $36\,(12)$ & $57\,(24)$ \\
$\siggas\,[\siggasunit]$ & $26\,(17)$ & $35\,(16)$ \\
\hline
\multicolumn{3}{l}{\emph{Same as the fiducial model, but using a beam-smearing cut of $\sigma_v < v_\mathrm{grad}$}} \vspace{0.05cm} \\
Mach number ($\mathcal{M}$) & $37\,(14)$ & $59\,(23)$ \\
$\siggas\,[\siggasunit]$ & $25\,(16)$ & $34\,(15)$ \\
\hline
\multicolumn{3}{l}{\emph{Same as fiducial model, but assuming the molecular velocity dispersion is equal to the H$\alpha$ velocity dispersion, $\sigma_v=\sigma_v(\mathrm{H}\alpha)$}} \vspace{0.05cm} \\
Mach number ($\mathcal{M}$) & $71\,(24)$ & $110\,(48)$ \\
$\siggas\,[\siggasunit]$ & $19\,(12)$ & $25\,(11)$ \\
\hline
\end{tabular}\label{tab:sum}
\begin{minipage}{\linewidth}
\vspace{0.1cm}
\textbf{Notes.} Note that the values in brackets denote the standard deviation (galaxy-to-galaxy variations) of each physical parameter; not the uncertainty in the parameter. Uncertainties are discussed in Sec.~\ref{sec:uncertainties}.
\end{minipage}
\end{table*}

Fig.~\ref{fig:derivation} suggests that the distributions of $\siggas$ and $\sigsfr$ are similar between the Star-forming and {\cas} galaxies. To quantify any statistical differences in $\siggas$ between these two sub-samples, we investigate the distribution functions of $\siggas$. Fig.~\ref{fig:gas_histograms} shows the histograms of $\siggas$. We see that $\siggas$ is enhanced in {\cas} galaxies compared to Star-forming galaxies. This difference in $\siggas$ between Star-forming and {\cas} type galaxies is primarily a consequence of the differences in $\sigsfr$, and secondarily a consequence of the differences in $\mach$ between the two classes, i.e., the dependences of $\siggas$ on $\sigsfr$ and $\mach$ (see Eqs.~\ref{eq:gas} and \ref{eq:gas_uncertainty}). Other dependences are realtively insignificant, such as the dependence on the assumed scale height of the galaxies. In this context, we have checked that any differences in the ellipticity distributions between the Star-forming and {\cas} galaxies are statistically insignificant.

The measured mean and standard deviation of $\sigsfr$, $\mach$ and $\siggas$ in the Star-forming and {\cas} galaxy samples are listed in Table~\ref{tab:sum} (the full list of physical parameters derived for each galaxy is provided in Table~\ref{tab:galaxies}). The SFR surface densities and Mach numbers of the Star-forming and {\cas} sample are $\sigsfr=0.054$ and $0.11\,\sigsfrunit$, and $\mach=36$ and $57$, respectively. The resulting average $\siggas$ values are $26$ and $35\,\siggasunit$ for Star-forming and {\cas} galaxies, respectively.

Table~\ref{tab:sum} further shows that changing the beam-smearing cutoff from the fiducial $\sigma_v < 2\,v_\mathrm{grad}$ to a less strict cutoff ($\sigma_v < v_\mathrm{grad}$) yields nearly identical results. Finally, the last two rows of Table~\ref{tab:sum} show that using the velocity dispersion of the ionized gas ($\sigma_v=\sigma_v(\mathrm{H}\alpha)$) instead of the approximate velocity dispersion of the molecular gas ($\sigma_v=\sigma_v(\mathrm{H}\alpha)/2$), reduces the derived $\siggas$ by $30\%$. Thus, even with the large uncertainties in $\sigma_v$ and hence in $\mach$ (see Sec.~\ref{sec:mach}), our final estimates in $\siggas$ can be considered accurate to within a factor 2.

\subsubsection{Mach number in Star-forming and {\cas} galaxies}

\begin{figure}
\centerline{\includegraphics[width=1.0\linewidth]{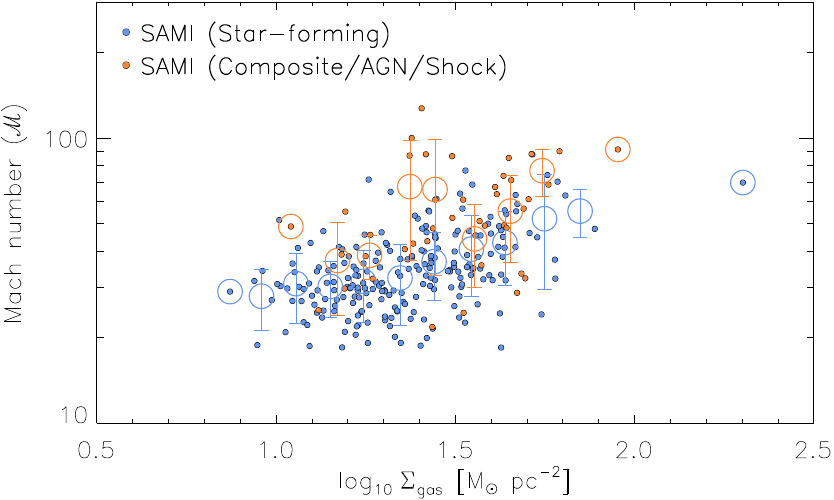}}
\caption{Turbulent Mach number ($\mach$) as a function of the derived $\siggas$ for Star-forming (blue) and {\cas} (orange) galaxies in our SAMI sample. The large open circles show the same data, but binned in steps of 0.1~dex in $\siggas$. The error bars show the standard deviation of the Mach number in each bin. Despite the significant variations in Mach number per individual $\siggas$ bin, {\cas} galaxies tend to have on average $\sim\!50\%$ higher $\mach$ than Star-forming galaxies.}
\label{fig:mach_gas}
\end{figure}

Here we investigate global differences in the gas kinematics between Star-forming and {\cas} galaxies in our SAMI sample. Fig.~\ref{fig:mach_gas} shows our measurements of the Mach number (c.f.~Sec.~\ref{sec:mach}) as a function of derived $\siggas$ for the two galaxy classes. We see that overall and also for fixed $\siggas$, {\cas} galaxies have higher Mach number by a factor of $\sim\!1.5$ compared to Star-forming galaxies. This may be a consequence of AGN and/or shocks raising the velocity dispersion over turbulence driven by pure star-formation feedback.

Fig.~\ref{fig:mach_gas} further reveals a significant scatter in Mach number for fixed $\siggas$, which is somewhat more pronounced in the case of {\cas} galaxies compared to purely Star-forming ones. This may indicate that different driving sources of the turbulence act together and possibly dominate at different times in different galaxies. Such driving sources can be divided into two main categories: i) stellar feedback (such as supernova explosions, stellar jets, and/or radiation pressure) and ii) galaxy dynamics (such as galactic shear, magneto-rotational instability, gravitational instabilities, and/or accretion onto the galaxy) \citep{FederrathEtAl2017iaus}. Our results here suggest that AGN feedback may be another important, potentially highly variable source of the turbulent gas velocity dispersion in galaxies.

\section{Comparison to previous $\sigsfr$ versus $\siggas$ relations} \label{sec:comp}

\begin{figure*}
\centerline{\includegraphics[width=0.85\linewidth]{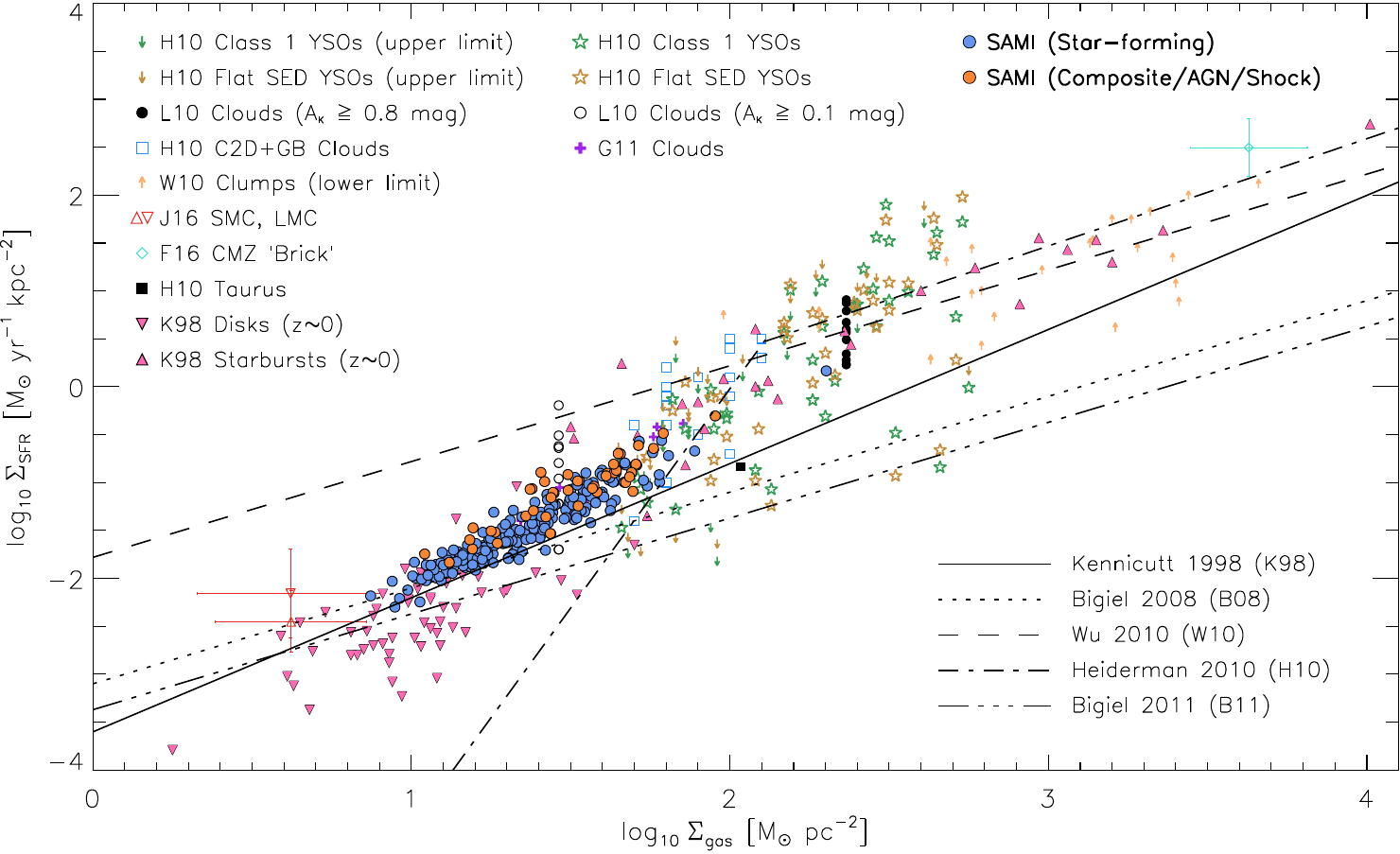}}
\caption{An enhancement of the right-hand panel of Fig.~\ref{fig:derivation}, but here we plot the individual Milky-Way clouds instead of the averages. Previous star formation relations from the literature are overlaid for comparison: K98 (solid line), \citet{BigielEtAl2008} (B08, dotted line), \citet{WuEtAl2010} (W10, dashed line), \citet{HeidermanEtAl2010} (H10, dash-dot line), and \citet{Bigiel:2011aa} (B11, dash-tripple-dot line).}
\label{fig:gas_comp}
\end{figure*}

Many studies in the literature have attempted to measure the correlation between $\sigsfr$ and $\siggas$ within different sets of data. However, there is no clear consensus on the coefficients and scaling exponents, due to the intrinsic scatter in the Kennicutt-Schmidt relation \citep[KDM12,][SFK15]{Federrath2013sflaw}. Some studies find breaks in the power-law relations, which can be interpreted as thresholds \citep{HeidermanEtAl2010}, while other studies do not find evidence for such thresholds \citep{K98,BigielEtAl2008,WuEtAl2010,Bigiel:2011aa}. Here we explore how our $\siggas$ predictions for the SAMI galaxies compare to the relations published in the literature.

\begin{figure*}
\centerline{\includegraphics[width=0.66\linewidth]{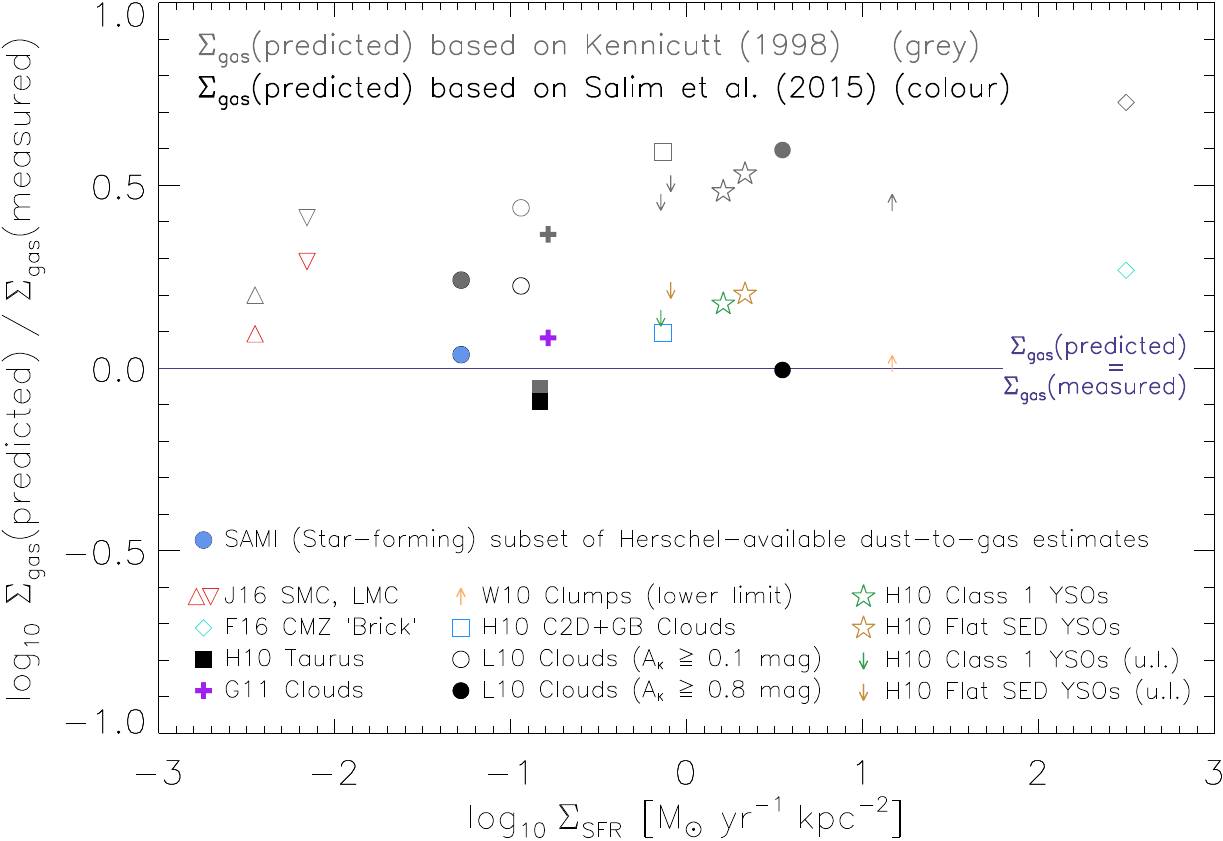}}
\caption{Logarithmic difference between $\siggas$(predicted) and $\siggas$(measured) as a function of $\sigsfr$ for the observational data shown in Fig.~\ref{fig:derivation} (except for the K98 galaxies, for which we currently do not have Mach number estimates). The grey data points show $\siggas$(predicted) based on inverting the K98 relation, while the coloured data points show the prediction based on inverting the SFK15 relation, i.e., the new method to estimate $\siggas$ from Eq.~(\ref{eq:gas}) developed here. The horizontal line shows $\siggas\mathrm{(predicted)}=\siggas\mathrm{(measured)}$. The SAMI data point (filled blue circle) is an average over 56 of our SAMI Star-forming galaxies for which \emph{Herschel} dust-to-gas estimates based on the method in \citet{GrovesEtAl2015} were available. We see that our new method based on the SFK15 relation provides a significantly more accurate prediction of $\siggas$ than inverting the K98 relation, with an average deviation of 0.12~dex (32\%) and 0.42~dex (160\%) for SFK15 and K98, respectively.}
\label{fig:sf_law_comp}
\end{figure*}

In Fig.~\ref{fig:gas_comp}, we show an enhanced version of the right-hand panel of Fig.~\ref{fig:derivation}, in order to compare our $\siggas$ estimates to previously derived star formation relations in the $\sigsfr$ versus $\siggas$ framework. The relations we investigate are described in K98, \citet{BigielEtAl2008} (B08), \citet{WuEtAl2010} (W10), \citet{HeidermanEtAl2010} (H10) and \citet{Bigiel:2011aa} (B11). These relations are shown as lines in Fig.~\ref{fig:gas_comp}. We see that different sets of data follow different relations, which often show significant deviations from one another.

The SAMI galaxies have higher $\sigsfr$ than described by the K98, B08, H10, or B11 relations, but lower $\sigsfr$ than described by the W10 relation. A power-law fit to all the SAMI galaxies yields a power-law exponent of $1.6\pm0.1$ instead of $1.4$ (K98); see Eq.~(\ref{eq:k98}).

In summary, we find that none of the previously proposed scaling relations of $\sigsfr$ as a function of $\siggas$ describes the entirety of the data well. The reason for this is that the SFR ($\sigsfr$) depends on more than just gas density ($\siggas$). Instead, star formation also strongly depends on the turbulence of the gas (Mach number and driving mode), the magnetic field, and on the virial parameter \citep[][SFK15]{KrumholzMcKee2005,PadoanNordlund2011,HennebelleChabrier2011,FederrathKlessen2012,Federrath2013sflaw,HennebelleChabrier2013,PadoanEtAl2014}. A complete understanding and prediction of star formation requires taking into account the dependences on these variables, in addition to gas (surface) density.

\section{Comparing $\siggas$ predictions by inverting star formation laws} \label{sec:k98vssfk15}

Here we compare the prediction of $\siggas$ based on inverting the K98 relation with the $\siggas$ prediction based on the SFK15 framework developed here. First, we note that a popular way of obtaining $\siggas$ estimates in the absence of a direct measurement of it, is to invert the K98 relation, i.e., to invert Eq.~(\ref{eq:k98}), which yields
\begin{equation} \label{eq:k98inverse}
\siggas{_\mathrm{,K98}}\,\left[\siggasunit\right] = \left(\frac{\sigsfr\,\left[\sigsfrunit\right]}{a}\right)^{1/n},
\end{equation}
with $a=(2.5\pm0.7)\!\times\!10^{-4}$ and $n=1.40\pm0.15$ (K98).

Here we derived an alternative way to estimate $\siggas$ from $\sigsfr$, which is given by Eq.~(\ref{eq:gas}), and contains additional dependences on the geometry ($L$), turbulence ($\mach$ and $b$), and on the magnetic field ($\beta$). For the $\siggas$ estimates of the SAMI galaxies here, we fixed $L$, $b$ and $\beta$ for simplicity, and only included the Mach number based on the measured velocity dispersion as an additional parameter to $\sigsfr$ (compared to the K98 relation, which depends on $\sigsfr$ only).

We now want to see how the $\siggas$ estimates based on our new relation (Eq.~\ref{eq:gas}) compare to inverting the K98 relation (Eq.~\ref{eq:k98inverse}). Fig.~\ref{fig:sf_law_comp} shows the direct comparison of the two as a function of $\sigsfr$. We plot the logarithmic difference of $\siggas$(predicted) and $\siggas$(measured) on the ordinate of Fig.~\ref{fig:sf_law_comp} for $\siggas$(predicted) based on K98 (Eq.~\ref{eq:k98inverse}) in grey and $\siggas$(predicted) based on Eq.~(\ref{eq:gas}) in colour. In addition to the observational data already shown in Figs.~\ref{fig:derivation} and~\ref{fig:gas_comp}, we add direct estimates of $\siggas$ for a subset of 56~Star-forming SAMI galaxies for which \emph{Herschel} $500\,\mu\mathrm{m}$ dust measurements were available, using the methods in \citet{GrovesEtAl2015}. A detailed description of how the dust emission was converted to $\siggas$ is provided in Appendix~\ref{app:herschel_dust_to_gas}.

In Fig.~\ref{fig:sf_law_comp} we see that our new method of estimating $\siggas$ from $\sigsfr$ given by Eq.~(\ref{eq:gas}) is significantly better than simply inverting the K98 relation, Eq.~(\ref{eq:k98inverse}). We find that our new relation provides $\siggas$ estimates with an average deviation of 0.12~dex (32\%), while inverting the K98 relation yields an average deviation from the true (measured) $\siggas$ by 0.42~dex (160\%). This shows that our method provides a significantly better $\siggas$ prediction from $\sigsfr$ than inverting the K98 relation. Our improved $\siggas$ estimate comes at the cost of requiring an estimate of the Mach number (velocity dispersion) as an additional parameter for the reconstruction (prediction) of $\siggas$. However, if $\sigsfr$ is obtained from H$\alpha$ (as for the SAMI galaxies analysed here), then we have shown that the velocity dispersion of H$\alpha$ can be used to estimate the Mach number (Sec.~\ref{sec:mach}).

Even better $\siggas$ predictions based on Eq.~(\ref{eq:gas}) are expected if the exact scale height $H$, the turbulence driving parameter $b$, and the magnetic field plasma $\beta$ are available from future observations and/or by combining different observational datasets.

\section{Conclusions} \label{sec:conc}

We presented a new method to estimate the molecular gas column density ($\siggas$) of a galaxy using only optical IFS data, by inverting the star formation relation derived in SFK15. Our method utilises observed values of $\sigsfr$ and velocity dispersion (here from H$\alpha$) as inputs and returns an estimate of the molecular $\siggas$. The derivation of our method is explained in detail in Sec.~\ref{sec:theory}, with the final result given by Eq.~(\ref{eq:gas}). We apply our new method to estimate $\siggas$ for Star-forming and {\cas} galaxies classified and observed in the SAMI Galaxy Survey.

Our main findings from this study are the following:

\begin{itemize}

\item From the range in \mbox{$\sigsfr=0.005$--$1.5\,\sigsfrunit$} and Mach number \mbox{$\mach=18$--$130$} measured for the SAMI galaxies, we predict \mbox{$\siggas=7$--$200\,\siggasunit$} in the star-forming regions of our SAMI galaxy sample, consisting of 260 galaxies in total. The predicted values of $\siggas$ are similar to those of unresolved low-redshift disc galaxies observed in K98. While the K98 galaxies required CO detections, here we estimate $\siggas$ solely based on H$\alpha$ emission lines.

\item We classify each galaxy in our sample as Star-forming or {\cas}. Based on the sample-averaged $\sigsfr=0.054$ and $0.11\,\sigsfrunit$, and $\mach=36$ and $57$ for Star-forming and {\cas} galaxies, respectively, we estimate $\siggas=26$ and $35\,\siggasunit$, respectively (see Table~\ref{tab:sum}). We therefore find that on average, the {\cas} galaxies have enhanced $\sigsfr$, $\mach$, and $\siggas$ by factors of $2.0$, $1.6$, and $1.3$, respectively, compared to the Star-forming SAMI galaxies (see Table~\ref{tab:sum}; for each individual SAMI galaxy, see Table~\ref{tab:galaxies}).

\item We discussed methods to account for finite spectral resolution and beam-smearing in Sec.~\ref{subsec:veldisp}. While the uncertainties are large in the velocity dispersion used to estimate the turbulent Mach number of the molecular gas (Sec.~\ref{sec:mach}), we show that the final estimate of $\siggas$ is accurate to within a factor of 2 (see Sec.~\ref{sec:uncertainties}).

\item We compare our new method of estimating $\siggas$ from $\sigsfr$ with a simple inversion of the K98 relation (Fig.~\ref{fig:sf_law_comp}). We find that our new method yields a significantly better estimate of $\siggas$ than inverting the K98 relation, with average deviations from the intrinsic $\siggas$ by 32\% for our new method, compared to average deviations of 160\% from inverting the K98 relation.

\end{itemize}

\section*{Acknowledgements}
We thank Mark Krumholz and the anonymous referee for their useful comments, which helped to improve this work.
CF acknowledges funding provided by the Australian Research Council's (ARC) Discovery Projects (grants~DP150104329 and~DP170100603).
DMS is supported by an Australian Government's New Colombo Plan scholarship. 
Support for AMM is provided by NASA through Hubble Fellowship grant \#HST-HF2-51377 awarded by the Space Telescope Science Institute, which is operated by the Association of Universities for Research in Astronomy, Inc., for NASA, under contract NAS5-26555. 
BAG gratefully acknowledges the support of the ARC as the recipient of a Future Fellowship (FT140101202). 
LJK gratefully acknowledges the support of an ARC Laureate Fellowship.
SB acknowledges the funding support from the ARC through a Future Fellowship (FT140101166).
SMC acknowledges the support of an ARC Future Fellowship (FT100100457).
NS acknowledges support of a University of Sydney Postdoctoral Research Fellowship.
The SAMI Galaxy Survey is based on observations made at the Anglo-Australian Telescope. 
The Sydney-AAO Multi-object Integral field spectrograph (SAMI) was developed jointly by the University of Sydney and the Australian Astronomical Observatory. The SAMI input catalogue is based on data taken from the Sloan Digital Sky Survey, the GAMA Survey and the VST ATLAS Survey. The SAMI Galaxy Survey is funded by the ARC Centre of Excellence for All-sky Astrophysics (CAASTRO), through project number CE110001020, and other participating institutions. The SAMI Galaxy Survey website is \url{http://sami-survey.org/}.

\appendix

\section{Online data} \label{app:online}

Table~\ref{tab:galaxies} lists the derived physical parameters of all SAMI galaxies analysed here. Listed are the first ten galaxies in each of our two galaxy classes (Star-forming and {\cas}). The complete table is available in the online version of the journal or upon request.

\begin{table*}
\caption{Spaxel-averaged physical parameters required to derive an estimate of the molecular gas surface density $\siggas$. Here only the first ten galaxies in each of our two samples classified as Star-forming or {\cas} are shown. The complete table is available in the online version of the journal.}
\def\arraystretch{1.0}
\setlength{\tabcolsep}{2.3pt}
\begin{tabular}{lccccccccccc}
\hline
GAMA & redshift & $\varepsilon$ & $N_\mathrm{spax}$ & $L_\mathrm{spax}$ & $\sigsfr$ & $\mathcal{M}$ & $\rho$ & $\mathrm{t_{ff}}$ & $\gastmff$ & $\gastff$ & $\siggas$ \\
ID   & & & & [$\mathrm{pc}$] & $\mathrm{[\sigsfrunit]}$ & & $[\mathrm{10^{-24}\,g\,cm^{-3}}]$ & [$\Myr$]& $\mathrm{[\sigsfrunit]}$ & $\mathrm{[\sigsfrunit]}$ & $\mathrm{[\siggasunit]}$ \\
\hline
\multicolumn{12}{c}{\emph{Star-forming classified galaxies}} \vspace{0.05cm} \\
$8353$ & $0.020$ & $0.30$ & $361$ & $200$ & $0.036\pm0.007$ & $28\pm19$ & $12\pm6\phantom{0}$ & $19\pm5\phantom{0}$ & $7.9\pm1.6$ & $1.3\pm0.7$ & $25\pm16$ \\
$8570$ & $0.021$ & $0.65$ & $10$ & $210$ & $0.0057\pm0.0011$ & $21\pm15$ & $2.5\pm1.3$ & $42\pm11$ & $1.3\pm0.3$ & $0.25\pm0.14$ & $11\pm7\phantom{0}$ \\
$9352$ & $0.024$ & $0.17$ & $40$ & $250$ & $0.065\pm0.013$ & $32\pm23$ & $18\pm9\phantom{0}$ & $16\pm4\phantom{0}$ & $14\pm3\phantom{0}$ & $2.1\pm1.2$ & $33\pm20$ \\
$15218$ & $0.026$ & $0.69$ & $29$ & $260$ & $0.0062\pm0.0012$ & $34\pm24$ & $1.9\pm1.0$ & $48\pm12$ & $1.4\pm0.3$ & $0.19\pm0.11$ & $9.2\pm5.7$ \\
$16026$ & $0.054$ & $0.49$ & $36$ & $520$ & $0.10\pm0.02$ & $77\pm54$ & $12\pm6\phantom{0}$ & $19\pm5\phantom{0}$ & $23\pm5\phantom{0}$ & $1.7\pm1.0$ & $34\pm21$ \\
$16294$ & $0.029$ & $0.30$ & $11$ & $290$ & $0.010\pm0.002$ & $23\pm16$ & $5.6\pm2.8$ & $28\pm7\phantom{0}$ & $2.2\pm0.4$ & $0.43\pm0.24$ & $12\pm7\phantom{0}$ \\
$22633$ & $0.070$ & $0.17$ & $289$ & $660$ & $0.065\pm0.013$ & $35\pm25$ & $18\pm9\phantom{0}$ & $16\pm4\phantom{0}$ & $14\pm3\phantom{0}$ & $2.0\pm1.1$ & $31\pm19$ \\
$22839$ & $0.039$ & $0.29$ & $17$ & $390$ & $0.010\pm0.002$ & $30\pm21$ & $4.9\pm2.5$ & $30\pm8\phantom{0}$ & $2.2\pm0.4$ & $0.34\pm0.19$ & $10\pm6\phantom{0}$ \\
$22932$ & $0.039$ & $0.29$ & $94$ & $390$ & $0.013\pm0.003$ & $25\pm18$ & $6.5\pm3.3$ & $26\pm7\phantom{0}$ & $2.9\pm0.6$ & $0.52\pm0.29$ & $14\pm8\phantom{0}$ \\
$23591$ & $0.025$ & $0.15$ & $17$ & $260$ & $0.078\pm0.016$ & $36\pm25$ & $20\pm10$ & $15\pm4\phantom{0}$ & $17\pm3\phantom{0}$ & $2.3\pm1.3$ & $35\pm21$ \\
\dots & \dots & \dots & \dots & \dots & \dots & \dots & \dots & \dots & \dots & \dots & \dots \\
\hline
\multicolumn{12}{c}{\emph{{\cas} classified galaxies}} \vspace{0.05cm} \\
$69740$ & $0.013$ & $0.45$ & $221$ & $140$ & $0.13\pm0.03$ & $63\pm45$ & $15\pm8\phantom{0}$ & $17\pm4\phantom{0}$ & $28\pm6\phantom{0}$ & $2.4\pm1.4$ & $41\pm26$ \\
$78531$ & $0.055$ & $0.23$ & $12$ & $530$ & $0.12\pm0.02$ & $86\pm61$ & $16\pm8\phantom{0}$ & $17\pm4\phantom{0}$ & $27\pm5\phantom{0}$ & $1.9\pm1.1$ & $31\pm19$ \\
$85416$ & $0.019$ & $0.51$ & $91$ & $200$ & $0.16\pm0.03$ & $61\pm43$ & $17\pm8\phantom{0}$ & $16\pm4\phantom{0}$ & $34\pm7\phantom{0}$ & $3.1\pm1.8$ & $51\pm31$ \\
$99349$ & $0.020$ & $0.63$ & $104$ & $200$ & $0.081\pm0.016$ & $32\pm23$ & $12\pm6\phantom{0}$ & $19\pm5\phantom{0}$ & $18\pm4\phantom{0}$ & $2.6\pm1.5$ & $50\pm31$ \\
$106376$ & $0.040$ & $0.21$ & $211$ & $400$ & $0.10\pm0.02$ & $29\pm20$ & $25\pm13$ & $13\pm3\phantom{0}$ & $22\pm4\phantom{0}$ & $3.5\pm2.0$ & $47\pm29$ \\
$106389$ & $0.040$ & $0.59$ & $21$ & $400$ & $0.080\pm0.016$ & $49\pm35$ & $11\pm5\phantom{0}$ & $20\pm5\phantom{0}$ & $18\pm4\phantom{0}$ & $1.9\pm1.1$ & $38\pm24$ \\
$144239$ & $0.018$ & $0.54$ & $204$ & $190$ & $0.044\pm0.009$ & $43\pm31$ & $8.2\pm4.1$ & $23\pm6\phantom{0}$ & $9.7\pm1.9$ & $1.1\pm0.6$ & $26\pm16$ \\
$144320$ & $0.052$ & $0.23$ & $40$ & $500$ & $0.079\pm0.016$ & $61\pm43$ & $14\pm7\phantom{0}$ & $18\pm4\phantom{0}$ & $17\pm3\phantom{0}$ & $1.6\pm0.9$ & $28\pm17$ \\
$204799$ & $0.017$ & $0.40$ & $59$ & $180$ & $0.23\pm0.05$ & $69\pm48$ & $23\pm12$ & $14\pm3\phantom{0}$ & $50\pm10$ & $4.2\pm2.4$ & $58\pm36$ \\
$210660$ & $0.017$ & $0.48$ & $59$ & $170$ & $0.029\pm0.006$ & $22\pm15$ & $9.6\pm4.9$ & $21\pm5\phantom{0}$ & $6.5\pm1.3$ & $1.3\pm0.7$ & $27\pm17$ \\
\dots & \dots & \dots & \dots & \dots & \dots & \dots & \dots & \dots & \dots & \dots & \dots \\
\hline
\end{tabular}
\begin{minipage}{\linewidth}
\vspace{0.1cm}
\textbf{Notes.} All galaxy parameters are based on a straight average over all valid spaxels and the uncertainties were propagated based on the average values. Column~1: GAMA ID. Column~2: redshift. Column~3: ellipticity. Column~4: number of valid spaxels for gas column density estimate. Column~5: linear size of each spaxel. Column~6: spaxel-averaged $\Sigma_{\mathrm{SFR}}$. Column~7: spaxel-averaged turbulent Mach number. Column~8: spaxel-averaged gas volume density, estimated based on Eq.~(\ref{eq:rho}). Column~9: spaxel-averaged freefall time based on Eq.~(\ref{eq:tff}). Column~10: spaxel-averaged multi-freefall gas consumption rate, $\gastmff$; Eq.~(\ref{eq:gastmff}). Column~11: spaxel-averaged single-freefall gas consumption rate, $\gastff$; Eq.~(\ref{eq:gastff}). Column~12: spaxel-averaged molecular gas surface density $\siggas$, estimated with Eq.~(\ref{eq:gas}).
\end{minipage}
\label{tab:galaxies}
\end{table*}

\section{\emph{Herschel} dust-to-gas estimates for SAMI} \label{app:herschel_dust_to_gas}

To provide an independent measure of $\siggas$ for our SAMI galaxy sample, we used the empirical relation determined by \citet{GrovesEtAl2015}, correlating the total gas mass of galaxies with their sub-mm dust luminosities. Using a sample of nearby galaxies, \citet{GrovesEtAl2015} found that the total (atomic$\,+\,$molecular) gas mass of galaxies ($M_\mathrm{gas,tot}$) could be determined within 0.12~dex using the monochromatic $500\,\mu\mathrm{m}$ luminosity ($L_{500}$), with
\begin{equation}
\log_{10}(M_\mathrm{gas,tot}/\msol)=28.5\,\log_{10}(L_{500}/L_\odot).
\end{equation}

To determine the sub-mm luminosity of the SAMI galaxies, we made use of the \emph{Herschel}-ATLAS survey \citep{EalesEtAl2010}, a wide 550~square~degrees infrared survey of the sky by the \emph{Herschel} Space Observatory, that covers the GAMA regions from which the SAMI Galaxy Survey sample arise. 
In particular, we cross-matched the 219~star-forming SAMI galaxies classified here against the single-entry source catalog from \emph{Herschel}-ATLAS Data Release~1 \citep{ValianteEtAl2016,BourneEtAl2016}.\footnote{Available at \url{http://www.h-atlas.org/public-data/download}} Of the 219~SAMI Star-forming galaxies, 128~have \emph{Herschel} detections. Of these, 56~have significant detections (signal-to-noise~$\!>\!3$) in the SPIRE $500\,\mu\mathrm{m}$ band.   

To convert the total gas mass to a gas surface density we required a surface area over which the infrared flux is emitted. Given the large beam size of the SPIRE $500\,\mu\mathrm{m}$ observations, the SAMI galaxies are unresolved. However, as can be seen in the radial profiles of the nearby galaxy sample used in \citet{GrovesEtAl2015} (in particular their Figure~7 and online figures), the highest surface brightness regions occur within half an optical radius \citep[$\sim\!0.5\,R_{25}$ or $1.8\,R_\mathrm{e}$ based on][]{WilliamsEtAl2010}, with most of the infrared luminosity (and molecular gas mass) occurring within this radius. \citet{GrovesEtAl2015} further find that at this radius, the atomic and molecular gas surface densities are about the same (the ratio of the total atomic and molecular gas masses inside $0.5\,R_{25}$ is also about unity). Based on those findings, we approximated the molecular gas mass within $0.5\,R_{25}$ with $0.5\,M_\mathrm{gas,tot}$. Therefore, we derive the molecular $\siggas$ through
\begin{equation}
\siggas=\frac{0.5\,M_\mathrm{gas,tot}}{\pi\left(1.8\,R_\mathrm{e}\right)^2 \left(1-\varepsilon\right)},
\end{equation}
where the effective radius $R_\mathrm{e}$, and ellipticity $\varepsilon$, of the SAMI galaxies are as derived in the GAMA survey \citep{DriverEtAl2011}.


\bsp	

\label{lastpage}

\end{document}